\documentclass[aps,prd,preprint,tightening,showpacs]{revtex4}
\usepackage{amsmath,amssymb,amsthm,graphicx}
\usepackage[mathscr]{eucal}
\usepackage[colorlinks,linkcolor=blue,anchorcolor=blue,citecolor=green]{hyperref}
\usepackage[dvipsnames,usenames]{color}
\usepackage{graphicx}
\usepackage{subfigure}
\usepackage{amsmath}
\newcommand{\no}{\nonumber}
\newcommand{\be}{\begin{equation}}
\newcommand{\ee}{\end{equation}}
\newcommand{\ba}{\begin{eqnarray}}
\newcommand{\ea}{\end{eqnarray}}
\usepackage{epsf,epsfig,graphics}
\usepackage{verbatim,color,ulem}
\begin{document}
\title{Gravitational time delay effects by Kerr and Kerr-Newman black holes in strong field limits }
\author{Tien Hsieh}
\author{Da-Shin Lee}
\email{dslee@gms.ndhu.edu.tw}
\author{Chi-Yong Lin}
\email{lcyong@gms.ndhu.edu.tw}
\affiliation{
Department of Physics, National Dong Hwa University, Hualien 97401, Taiwan, Republic of China}
\date{\today}

\begin{abstract}
We study the time delay between two relativistic images due to strong gravitational lensing of the light rays caused by the Kerr and Kerr-Newman black holes.
{The trajectories of the light rays are restricted on the equatorial plane.}
Using the known form of the deflection angle in the strong deflection limit (SDL) allows us to analytically develop the formalism for the travel time of the light  from the distant source winding around the black hole several times and reaching the observer.
We find that the black hole with higher mass or with spin of the extreme black hole potentially have higher time delay.
%
The effect of the charge of the black hole  enhances the time delay between the images lying on the opposite side of  the optical axis resulting from the light rays when one light ray is in the direct orbit and the other is in the retrograde orbit. In contrary, when both light rays  travel  along either direct or retrograde  orbits giving the images on the same side of the optical axis, the charge effect reduces  the time delay between them.
%
We then examine the time delay observations due to the galactic and supermassive black holes respectively.

\end{abstract}


\maketitle

\section{Introduction}
{ General Relativity (GR) describes how mass concentrations distort the space around them where the gravitational lens can occur when massive astrophysical objects, such as stars, galaxies and clusters of galaxies,
create a gravitational field to distort the light that passes by these objects. \cite{Dyson_1920,MIS_1973,Weinberg_1973,Hartle_2003}.
When the gravitational field is weak and the deflection angles of the light rays are small, the weak field approximation is adequate to describe the weak lensing.
\cite{Schneider_1992,Refsdal_1994}.
Nevertheless, when the light rays enter the strong gravitational field near the black hole, the deflection angles can be arbitrary large so that the light rays can circle around the black hole multiple times and then reach the observers. The light deflection in strong gravitational field of  Schwarzschild black holes was  studies  several decades ago  by
Darwin \cite{Darwin_1959}, and was reexamined in \cite{Luminet_1979,Ohanian_1987,Nemiro_1993}.
}
Since then, there have been significant theoretical efforts  for understanding lensing  from strong field perspectives, to cite a few \cite{Virbhadra_2000,Frittelli_2000,Bozza_2001,Bozza_2002,
Bozza_2003,Eiroa_2002,Bozza_2007,Iyer_2007,Tsukamoto_2017a,Tsukamoto_2017b,Gralla_2020a,Gralla_2020b}.
Also, inspired by  the first image of  the black hole  captured by the Event Horizon Telescope \cite{EHT1,EHT2,EHT3}, the exploration of the properties of  the isolated dim black hole can rely on the lensing effects of the light rays from the distant sources.

\begin{figure}[htp]
\begin{center}
\includegraphics[width=15cm]{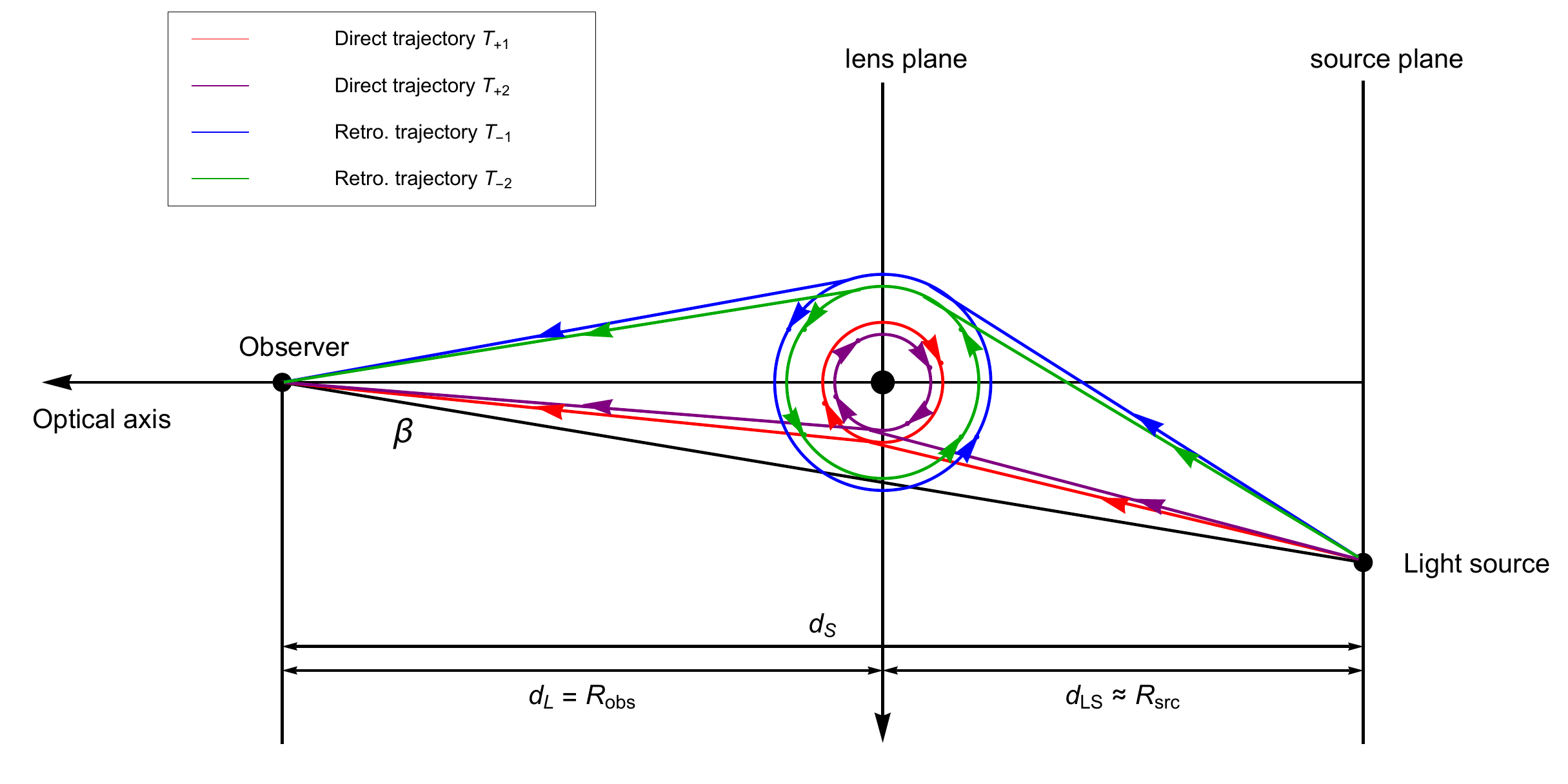}
\caption{
{
          The Kerr or the Kerr-Newman black hole is assumed to has angular momentum of the clockwise rotation.
          In the SDL, the light rays  from the distant  source approach the black hole,
            circle around it multiple times  along a direct orbit (red or purple line) or a retrograde orbit (blue or green line), and then reach the observer.  }}
        \label{fig:arch_02}
    \end{center}
\end{figure}

  For the general spherically symmetric and static spacetime, the deflection angle $\hat{\alpha}(b)$ of light rays { on the equatorial plane} in the limit of $b\to b_{c}$
can be approximated in the following simple form   \cite{Bozza_2001,Bozza_2002,Bozza_2003,Tsukamoto_2017a,Tsukamoto_2017b},
\begin{equation} \label{hatalpha_as}
\hat{\alpha}(b)\approx-\bar{a}\log{\left(\frac{b}{b_{c}}-1\right)}+\bar{b}+\mathcal{O}(b-b_{c}) \log(b-b_{c}))
\end{equation}
with the two parameters $\bar a$ and $\bar b$ depending on black hole's parameters.
%
{Recently in \cite{Gralla_2020a,Gralla_2020b} Gralla and  Lupsasca have provided the complete set of null geodesics of the Kerr exterior beyond the equatorial plane, with which, the analytical expressions for the deflection angles and the time delays are achieved. }
In \cite{Hsieh_2021}, we extended the works of \cite{Bozza_2002,Tsukamoto_2017a} and find the analytic form of $\bar a$ and $\bar b$  for   Kerr and Kerr-Newman black holes, respectively, with the analytical closed-form expressions of the deflection angles obtained in \cite{Iyer_2009} and \cite{Hsiao_2020}.
In \cite{Hsieh_2021}, we have examined the lensing effects due to the  black hole  when the light rays from the source circle around the black hole multiple times in the strong deflection limit (SDL)  along a direct orbit  or  a retrograde orbit, giving two sets of the relativistic images as illustrated in Fig.(\ref{fig:arch_02}).

In strong gravitational lensing phenomena, the apparent angles of the images and the time delay between relativistic images are  the most important observable that can be widely used to reveal information about the lens
from the resulting effects of the light rays \cite{Gralla_2020a,Gralla_2020b,Bozza_2004,Keeton_2005,Keeton_2006,Virbhadra_2008,Virbhadra_2009,Lu_2016}.
According to  \cite{Bozza_2004},  Bozza  developed the following formula of the travel time of the light on the equatorial plane of the Kerr black holes
%
\begin{equation}\label{travel_time_SDL}
T(b)= -2 \tilde{a} \log{\left( \frac{b}{b_{c}}-1 \right)}+\tilde{b}(R_{src}) +\tilde{b}(R_{obs})
+\mathcal{O}\left( (b-b_{c})\log(b-b_{c})   \right)
\end{equation}
in the strong field limit
by introducing another two parameters $\tilde a$ and $\tilde b$. In (\ref{travel_time_SDL}),  $\tilde a$ just depends on the black hole's parameters whereas $\tilde b$ not only depends on the black hole's parameters but also has the dependence of the distance of the source $R_{src}$ and the observer $R_{obs}$ from the lens.
{Extensive studies for the light rays travelling on the more general  nonequatorial plane in the Kerr spacetime can be found in \cite{Gralla_2020a,Gralla_2020b}.}
Following \cite{Bozza_2004}, we will work on the time delay effects due to  the Kerr and Kerr-Newman black holes \cite{Liu_2017,Jiang_2018,Kraniotis_2014}, the two important solutions resulting from  the Einstein equations plus the Maxwell equations to incorporate the charge of the black hole \cite{MIS_1973,adamo2016kerrnewman}.
%
In this work, we  will explicitly find out the analytical expressions of the coefficients $\tilde a$ and $\tilde b$ with which to extract interesting features of the time delay observations when the light ray are on the equatorial plane.

{In the case of the Kerr black holes, our analytical results will be compared with \cite{Bozza_2004} given numerically.
In particular the coefficient $\tilde a$ of the logarithmic term, which becomes dominant in the SDL as $
b \rightarrow b_c$ in   (\ref{travel_time_SDL}) will show the consistency with  \cite{Gralla_2020b} in the limit of the light rays lying on the equatorial plane. Additionally, the coefficient $\bar a$ in the SDL deflection angle (\ref{hatalpha_as}) in (\ref{abar_k})   in our previous work \cite{Hsieh_2021} can be further simplified in the form to compare  with \cite{Gralla_2020b} again, restricting on the equatorial plane.
In particular, the relation (\ref{ab_relation}), which was discovered in the  spherically symmetric and static spacetime \cite{Bozza_2004}, still holds true in the Kerr and Kerr-Newman black holes.
 The relation (\ref{ab_relation}) provides better  understanding of the time delay  effects between two relativistic images  in the formulas in (\ref{delta_T_s}) and (\ref{delta_T_o}) in a geometric way, which are determined by  the critical impact parameter $b_c$ forming the light sphere.
Apart from the coefficient $\tilde a$, the coefficient $\tilde b$ is explicitly determined by the dependence not only of the black hole's parameters but also of the distances of the source and the observer from the black holes, $R_{src}$ and $R_{obs}$ respectively. As in \cite{Gralla_2020b}, when considering the time delay between two relativistic images, the dependence of $R_{src}$ and $R_{obs}$ cancels out resulting in the dependence of the black hole's parameters only.
We then extend the studies to the Kerr-Newman black holes to see how the charge of the black holes influences the time delay effects.
In addition to the time delay, in \cite{Stefanov_2010,Raffaelli_2016}, the applications of the coefficients $\bar a$ and $\bar b$ with the related  $\tilde a$ and $\tilde b$  are proposed to give an alternative way for the measurement of  quasinormal frequencies of the black holes, which play importance roles in gravitational-wave astrophysics. }

Layout of the paper is as follows. In Sec.II, we first develop the formalism of the travel time of the light  from the source circulating around the black hole and reaching the observer.
Then we derive the analytic form of $\tilde a$ and $\tilde b$ in (\ref{travel_time_SDL}) for the cases of Kerr and Kerr-Newman black holes, respectively, and check the consistency with the known results from taking the proper limits of the black holes's parameters.
In addition, we will find their approximate expressions in the limits as the spin of the black hole is small $a\to 0$, as well as the spin reaches the extreme value $a\to M$ ($a\to \sqrt{M^2-Q^2}$) of  the extreme Kerr (Kerr-Newman) black holes.
In Sec. III, the analytical expressions of $\tilde a$ and $\tilde b$  are then applied to compute the
time delay between relativistic images obtaining their general features.
The approximately analytical forms of $\tilde a$ and $\tilde b$ are very useful to interpret the time delay behavior in the appropriates limits.
The effects of angular momentum and charge of the black holes to the time delay observations will be summarized in the closing section.
Additionally, the closed-form expression of the deflection angle due to the Kerr and/or the Kerr-Newman black holes is summarized in Appendix A.
%
In particular, we present the results of the radius of the innermost circular motion of light rays as well as the associated critical impact parameters as a function of the black hole's parameters.  These will  serve as the important inputs to find the values of the coefficients $\bar a$ and $\bar b$ in the SDL deflection angle.
In Appendix B, we provide useful integral formulas to find the coefficients $\tilde a$ and $\tilde b$.
In this paper, we have used $c=G=1$ units, unless otherwise specified.

\section{Time delay due to black holes in the strong field limit}

We consider nonspherically symmetric spacetimes of the Kerr and Kerr-Newman metrics to obtain the travel time $T(b)$ of light  from the  source to the observer for a given impact parameter $b$.
In the SDL, as $b\to b_{c}$, $T(b)$ can be approximated in the form (\ref{travel_time})
with two parameters $\tilde{a}$ and $\tilde{b}$ as a function of the black hole's parameters and the distances of the source and the observer from the black hole.
In what follows, we will discuss these two kinds of the black holes separately.

\subsection{Kerr black holes}

The spacetime outside of the horizon of the Kerr black hole with the gravitational mass $M$ and  angular momentum per unit mass $a=J/M$ is described by the line element below
\ba
 {ds}^2
&=& g_{\mu\nu} dx^\mu dx^\nu  \nonumber \\
&=& -\frac{ \left(\Delta -a^2 \sin^2\theta \right)}{\Sigma } {dt}^2   - \frac{ a   \sin ^2\theta \left(2 M
   r\right)}{\Sigma } ({dt}{d\phi+ d\phi dt)}  \nonumber\\
&&
+\frac{\Sigma}{\Delta} dr^2 +\Sigma{\, d\theta}^2 +\frac{ \sin ^2 \theta}{\Sigma} \left(( r^2+a^2)^2 -a^2 \Delta  \sin
   ^2\theta \right) {d\phi}^2 \,  \label{Kerr_metric}
   \ea
with
\be \label{Delta_k}
\Sigma=r^2+a^2\cos^2\theta \, , \quad  \Delta=r^2+a^2-2 M r\,.
\ee
Solving $\Delta(r)=0$ gives the outer (inner)  event horizon $r_+$ ($r_-$) as
\be \label{rpm_k}
r_{\pm}=M\pm\sqrt{M^2-a^2}\,
\ee
with the condition $M^2 > a^2$.
Notice that  although the inner event horizon  $r_-$ is introduced, here we consider the light rays traveling outside the outer horizon.
Due to the fact that the metric component functions are independent on $t$ and $\phi$, and thus the metric itself is both stationary and axial-symmetric.
Together with 4-velocity of light defined in terms of the affine parameter, the  conserved quantities of light ray's energy and azimuthal angular momentum along a geodesic obeying $ u^\mu u_\mu=0$  can be obtained from the  Killing vectors
$ \varepsilon  \equiv -\xi_{(t)}^\mu u_\mu $ and $\ell  \equiv \xi_{\phi}^\mu u_\mu $,
where  the associated Killing vectors
$\xi_{(t)}^\mu$ and $\xi_{(\phi)}^\mu$ are $
\xi_{(t)}^\mu=\delta_t^\mu \, ,  \xi_{\phi}^\mu=\delta_\phi^\mu \,.$
The light  can be traversing along the direction of the black hole rotation or opposite to it with the impact parameter  defined as
 \begin{align}
 b_s=s\left|\frac{\ell}{\varepsilon}\right|\equiv s\,b \; ,
 \end{align}
where $s=\text{Sign$(\ell/\varepsilon)$}$ and $b$ is the positive magnitude.
The parameter $s=+1$ for $b_s>0$ will be referred to as direct orbits, and those with $s=-1$ for $b_s<0$ as retrograde orbits (see Fig.(\ref{fig:arch_02}) for the sign convention).
In this paper,  the light rays are restricted to traveling on the equatorial plane of the black hole, where $\theta={\pi}/{2}$, and $\dot{\theta}=0$.
In the end, their dynamics  can be summarized into the equation of motion along the radial direction  in the form  \cite{Hsiao_2020}
 \begin{align} \label{r_eq}
 \frac{1}{b^2} & =\frac{\dot{r}^2}{\ell^2}+W_\text{eff}(r)\; ,
 \end{align}
where the effective potential function $W_\text{eff}$ is found as
\begin{align}
W_\text{eff}(r)=\frac{1}{r^2}\left[1-\frac{a^2}{b^2}-\frac{2M }{r}\left(1-\frac{a}{b_s}\right)^2\right] \, . \label{eq:Weff_k}
\end{align}

In the SDL, as seen in Fig (\ref{fig:arch_02}), the light rays  starting in the asymptotic region approach the black hole, wind around it multiple times, and then reach  the observer in the asymptotic region.
%
{During the journey, the light rays will reach the turning point, the  closest approach distance to a black hole $r_0$, and in particular the smallest radius ${r_{sc}}$ of the innermost trajectories of light can be given by the turning point $r_0$  located at the maximum of $W_\text{eff}(r)$, satisfying the follow equations
\begin{equation}\label{rsc}
\left.\frac{\dot{r}^2}{\ell^2}\right|_{r=r_0}=\frac{1}{b^2}-W_\text{eff}(r_0)=0 \,, \quad\quad
 \left.\frac{d\,W_\text{eff}(r)}{dr}\right|_{r=r_{sc} }  =0  \, .
 \end{equation}}
The solutions are reviewed in Appendix A.
In fact,  the values of $ b_{sc} $ and $r_{sc}$ are the key inputs to determine the features of the deflection angle of the distant light sources  due to the strong  gravitational lensing effects.
Given the effective potential  $W_\text{eff}$ in (\ref{eq:Weff_k}), the nonzero spin of the black holes induces more repulsive  effects  to the light in the direct  orbits than those in the retrograde orbits due to  the $1/r^3$ term.
%
Near horizon the repulsive  effects turn out effectively to prevent the light  in the direct orbits from collapsing into the event horizon.
As a result, this shifts the innermost circular trajectories  of the light rays  toward the black hole with the smaller critical impact parameter $b_{+c}$ than  $b_{-c}$ in the retrograde orbits as shown in \cite{Iyer_2009,Hsiao_2020}.
As such, when $a$ increases, the impact parameter $b_{+c}$ decreases whereas $ b_{-c}$ increases instead.
In addition, the previous work \cite{Hsieh_2021} also shows that the deflection angle $\hat{\alpha}(b)$ of light  for a given impact parameter $b$, which  in the SDL as $b\to b_{c}$, can be approximated in the form
$$
\hat{\alpha}(b)\approx-\bar{a}\log{\left(\frac{b}{b_{c}}-1\right)}+\bar{b}+O(b-b_{c}) \log(b-b_{c}))\;,
$$
where the two parameters $\bar a$ and $\bar b$ are obtained in terms of black hole's parameters (See brief review in Appendix A).

Now consider  the travel time of the light   from the source to the observer, which pass by the black hole with the closest approach distance $r_0$  as
\begin{equation}\label{eq:rdot_Weff}
T(r_0)= \int_{R_{src}}^{r_0} \Big\vert \frac{dt}{dr} \Big\vert dr  +  \int_{r_0}^{R_{obs}} \Big\vert \frac{dt}{dr} \Big\vert dr\,.
\end{equation}
The parameters $R_{src}$ and $R_{obs}$ are the distance between the black hole and the source or observer, respectively.
 Following \cite{Hsiao_2020}, we can rewrite the equation of motion $dt/dr$ in (\ref{rsc}) by introducing a new variable $z$ defined by $z=1-r_0/r$ as
\begin{equation}\label{dt/dz_k}
\frac{dt}{dz}=\frac{r_0}{(1-z)^2}  \frac{-a \frac{2M}{r_0} (1-\frac{a}{b}) (1-z)^3 + \frac{r_0^2}{b} + \frac{a^2}{b}(1-z)^2}{1-\frac{2M}{r_0}(1-z)+\frac{a^2}{r_0^2}(1-z)^2}  \frac{1}{\sqrt{B(z,r_0)}}\;,
\end{equation}
where the function $B(z, r_0)$ has the trinomial form in $z$
\begin{equation}
B(z,r_0)=c_1(r_0)z+c_2(r_0)z^2+c_3(r_0)z^3 \,
\end{equation}
 with the coefficients
\begin{eqnarray}
c_1(r_0)&=&-6Mr_0\left(1-\frac{a}{b_s}\right)^2+2r_0^2\left(1-\frac{a^2}{b^2}\right)\,,\label{c1_k}\\
c_2(r_0)&=&6Mr_0\left(1-\frac{a}{b_s}\right)^2-r_0^2\left(1-\frac{a^2}{b^2}\right)\, ,\label{c2_k}\\
c_3(r_0)&=&-2Mr_0\left(1-\frac{a}{b_s}\right)^2\, .\label{c3_k}
\end{eqnarray}
Then, $R_{src}$ and $R_{obs}$ can be expressed by $z$ as, $Z_{src}=1-r_0/R_{src}\frac{}{}$ and $Z_{obs}=1-r_0/R_{obs}\frac{}{}$, respectively.
The observer and the source are supposedly very far from the black holes, $R_{src},R_{obs} \gg r_0$, then both $Z_{src},  Z_{obs} \to 1$.

We rewrite part of the integrand in (\ref{dt/dz_k}) by the partial fraction,
\begin{equation}
\begin{split}
&\frac{1}{(1-z)^2}  \frac{-a \frac{2M}{r_0} (1-\frac{a}{b}) (1-z)^3 + \frac{r_0^2}{b} + \frac{a^2}{b}(1-z)^2}{1-\frac{2M}{r_0}(1-z)+\frac{a^2}{r_0^2}(1-z)^2} \\
& = \frac{r_0^2}{a^2} \frac{1}{b}  \left[  \frac{\tilde C_-}{z-z_-}+\frac{\tilde C_+}{z-z_+}+ \frac{r_0^2}{(1-z)^2 (z-z_-) (z-z_+)}  \right]
\end{split}
\end{equation}
The roots and the corresponding coefficients $\tilde C_-$, $\tilde C_+$ in the Kerr case are
\begin{eqnarray} \label{C_tilde_pm_k}
z_-&=&1-\frac{r_0r_-}{a^2}\;,\\
z_+&=&1-\frac{r_0r_+}{a^2}\;,\\
\tilde C_-&=& \frac{-2M a br_- +a^4 +2Ma^2 r_-}{2r_0 \sqrt{M^2-a^2}}\, ,\\
\tilde C_+&=& \frac{2M a br_+ -a^4 -2Ma^2 r_+}{2r_0 \sqrt{M^2-a^2}}\, ,
\end{eqnarray}
with $r_+$ ($r_-$) being the outer (inner) horizon of a Kerr black hole defined in (\ref{rpm_k}).
Notice that $z_-$, $z_+ \le 0$, for  all spin $a$.

Then the travel time of the light can be calculated as a function of the closest approach distance $r_0$ from (\ref{dt/dz_k}) giving
\begin{equation}\label{travel_time}
T(r_0) = I (Z_{src},r_0) +  I (Z_{obs},r_0) \, , \quad I(Z,r_0)= \int_0^Z f(z,r_0) dz \, .
\end{equation}
In the SDL,  when $r_0 \to r_{sc}$,  $c_1(r_0) \to 0$ in (\ref{c1_k}) obtained from (\ref{rsc}), the integral  $f(z,r_0) \rightarrow 1/z\frac{}{}$ for small $z$ will lead to the logarithmic divergence as  $r_0 \to r_{sc}$.
Let us now define a new function $f_D(z,r_0)$
\begin{equation}\label{f_D_k}
f_D(z,r_0)=\frac{r_0^3}{a^2 b}\left[ \frac{\tilde C_-}{z-z_-}+\frac{\tilde C_+}{z-z_+}+ \frac{r_0^2}{(1-z)^2 (z-z_-) (z-z_+)} \right] \frac{1}{\sqrt{c_1(r_0)z+c_2(r_0)z^2}}\;,
\end{equation}
and also $f_R(z,r_0)=f(z,r_0)-f_D(z,r_0)$. The integral of $f_R$ over $z$ is thus finite given by (\ref{I_R_k}) in Appendix B and
contributes part of $\tilde b$ in (\ref{travel_time_SDL}), denoted by $\tilde b_R$.
The divergent part due to an integral of the function $f_D(z,r_0)$ over $z$ can be obtained from the result of (\ref{I_D_k}), giving not only to $\tilde{a}$ for the logarithmic divergence but also contributing partly to the formula of $\tilde b$ for the regular part in (\ref{travel_time_SDL}), denoted by $\tilde b_D$.
{The possible divergence means  that the light rays spend long time in circling around the black hole as the impact parameter is near its critical value, $b \to b_{sc}$.}
Notice that in the SDL, all the coefficients $c_1, c_2$ and $c_3$  in (\ref{c1_k})-(\ref{c3_k}) will be evaluated at $r_0=r_{sc}$.
However, since $c_1(r_{sc})\equiv c_{1sc}=0$, we need the expansion of the coefficient $c_1(r_0)$ given by (\ref{c1_k}) to the next order, namely
\begin{equation} \label{c1_SDL}
c_1(r_0)=c_{1 sc}' (r_0-r_{sc})+\mathcal{O}(r_0-r_{sc})^2 \, .
\end{equation}
The subscript  ``{\it sc}'' denotes evaluating the function at $r=r_{sc}$. The prime means the derivative with respect to $r_0$.
Using $c_{1sc}=0$ in (\ref{c1_k}) leads to
\begin{equation}\label{c2c3_k}
 c_{3sc}=-\frac{2}{3}c_{2sc}\;.
 \end{equation}
Furthermore, one can write  $c_1(r_0)$  in terms of $b-b_{sc}$. To achieve it,  the expansion of
the impact parameter $b(r_0)$ at $r_{sc}$ reads
\begin{equation} \label{b_SDL}
b(r_0)=b_{sc}+\frac{b_{sc}''}{2!}(r_0-r_{sc})^2+\mathcal{O}(r_0-r_{sc})^3 \, .
\end{equation}
Combining  (\ref{c1_SDL}) and (\ref{b_SDL}) gives
\begin{equation} \label{c_b_SDL}
\lim_{r_0\to r_{sc}} c_1(r_0)=\lim_{b\to b_{sc}}c_{1 sc}'\sqrt{\frac{2b_{sc}}{b_{sc}''}}\left(\frac{b}{b_{sc}}-1\right)^{1/2}\;.
\end{equation}
%
Thus, the straightforward calculations give the coefficients $\tilde{a}$ and $\tilde{b}=\tilde b_R+\tilde b_D$ in the key formula (\ref{travel_time_SDL}) as
\be \label{tilde_a_k_f}
\tilde{a}= \frac{r_{sc}^3}{2 b_{sc} \sqrt{c_{2sc}}}\left[\frac{2M+r_{sc}}{2r_{sc}}+
  \frac{\tilde C_{- sc}}{r_{sc}r_{-}-a^2}
+\frac{a^6}{r_-^2 (r_{sc}r_- -a^2) r_{sc} (r_+-r_-)}\right]   +(- \leftrightarrow +)\; ,
\ee
which
{can be further simplified by
with (\ref{light_sphere_k}) and (\ref{bsc_k2}) to replacing $b_{sc}$ and  $a \sqrt{M r_{sc}}$  by $r_{sc}$ as
\begin{equation}\label{tilde_a_k}
\tilde a = \frac{1}{2} \frac{r_{sc}}{\sqrt{3}} \left( \frac{r_{sc}+3M}{r_{sc}-M} \right)  \,
\end{equation}
}
and
\begin{equation}\label{tilde_b_k}
\begin{split}
\tilde{b}=&\frac{1}{2}\tilde{a} \log{\left( \frac{8c_{2sc}^2 Z^2  b_{sc}''}{c_{1sc}'^2 b_{sc}} \right)} \\
&+ \frac{r_{sc}^3}{2 b_{sc}\sqrt{c_{2sc}}}
\left[ \frac{\sqrt{3}\sqrt{3-2Z}}{1-Z}+ 2 \sqrt{3}\frac{M}{r_{sc}} \log{\left(  \frac{2-Z+\sqrt{3-2Z}}{2+\sqrt{3}} \frac{1}{1-Z}\right)} \right. \\
&\left. \quad\quad \quad \quad\quad\quad \quad\quad -3-  \frac{r_{sc}^2}{(a^2-2M r_{sc}+r_{sc}^2)} \log\tilde{N}(Z,a)  \right]  \\
& +\frac{r_{sc}^3 \tilde C_-}{b_{sc}\sqrt{c_{2sc}}(a^2-r_{sc}r_-)} \left[ \log {\tilde N}(Z,a) - \frac{\sqrt{3}a}{\sqrt{a^2+2r_{sc}r_-}}
\log \tilde{M}(Z,a,r_{-}) \right]\\
&-\frac{r_{sc}^2}{b_{sc}\sqrt{c_{2sc}}}  \frac{\sqrt{3}\, a^7}{r_-^2 (a^2-r_{sc}r_-)\sqrt{a^2+2r_{sc}r_-}(r_+-r_-)} \log{ \tilde{M}(Z,a,r_{-})}+(- \leftrightarrow +)\;,
\end{split}
\end{equation}
where
\begin{equation}
\tilde{M} (Z,a,r_{\pm})=\frac{a^2-r_{sc}r_-}{a^2-r_{sc}r_--Za^2}
\frac{(2-Z)a^2+r_{sc}r_{\pm}+a\sqrt{3-2Z}\sqrt{a^2+2r_{sc}r_{\pm}}}{2a^2+r_{sc}r_-+a\sqrt{3}\sqrt{a^2+2r_{sc}r_{\pm}}}\; ,
\end{equation}
\begin{equation}
\tilde{N}(Z,a)= \frac{3-Z+\sqrt{3}\sqrt{3-2Z}}{6} \; .
\end{equation}
This is one of the main results in this paper. The analytical expressions of $\tilde a$ and $\tilde b$ in the SDL travel time of the light  around the black hole can be applied  to compute the time delay between two relativistic images. Through comparing with observations can provide information on the black hole's parameters \cite{Bozza_2004}.

Now let us compare with the some known results on the Schwarzschild black holes in \cite{Bozza_2004}.
In the limit of $a \to 0$,
$ r_+ \rightarrow 2M$, $ r_- \to a^2/2M$,
$ \tilde C_{+c} \to a$,
$ \tilde C_{-c}\to a^3$, and  $c_2 \to r_{c}^2$ using $c_{1sc}=0$.
Since  $b_{c}''\to\sqrt{3}/M$ and $r_{c}=3M$, the coefficient $\tilde a$ in (\ref{tilde_a_k}) can be simplified to
 \begin{equation}
 \tilde{a}=\frac{3\sqrt{3}}{2} M.
 \end{equation}
Together with $\bar a=1$ in the expression of the SDL deflection angle in (\ref{hatalpha_as}) gives the expected relation
  \begin{equation} \label{ab_relation}
 \frac{2 \tilde{a}}{\bar{a}}=b_{c}
 \end{equation}
for the spherically symmetric black holes.
%
%
{Next, we will compare with \cite{Gralla_2020b} when the light rays are on the equatorial plane. The coefficient $\tilde a$ can also be obtained by integrating  the $1/z$ dependence in (\ref{dt/dz_k}) over $z$, giving \cite{Gralla_2020b}
\begin{equation}\label{tilde_a_k_Gr}
\tilde a = \frac{r_{sc}}{4 \sqrt{\chi_{sc}}} \left( \frac{r_{sc}+3M}{r_{sc}-M} \right)  \,.
\end{equation}
The notation $\chi_{sc}$ is defined below and simplified to a number through the light sphere formula (\ref{light_sphere_k})
\begin{equation}\label{chi}
\chi_{sc} = 1-\frac{M \Delta(r_{sc})}{r_{sc}(r_{sc}-M)^2} = \frac{3}{4}\,,
\end{equation}
with $\Delta$  defined in (\ref{Delta_k}) for the Kerr cases.
Also, in \cite{Gralla_2020b} {the coefficient $\bar a$} is found to be
\begin{equation}\label{bar_a_k_Gr}
 \bar a = \left[ a \left( \frac{r_{sc}+M}{r_{sc}-M} \right) +a-\frac{2a r^2_{sc}}{ r_{sc}^2- 3 M r_{sc}} \right] \frac{1}{2 r_{sc} \sqrt{\chi_{sc}}}\,
\end{equation}
with again $\chi_{sc}=3/4$.
Through the same strategy of simplifying the coefficient $\tilde a$, the coefficient $\bar a$ in (\ref
{abar_k_f}) also has the simple version given by
\begin{equation}\label{abar_k}
\bar a = \frac{2M r_{sc}}{(r_{sc}-M) \sqrt{3 M r_{sc}}}  \,.
\end{equation}
Using the straightforward algebra and again (\ref{light_sphere_k}) one can show the coefficients $\tilde a$ and $\bar a$ in (\ref{tilde_a_k}) and (\ref{abar_k}) are the same as
those in (\ref{tilde_a_k_Gr}) and (\ref{bar_a_k_Gr}) obtained in \cite{Gralla_2020b}, which also obey the relation (\ref{ab_relation}).
 It is worthwhile to mention that the relation (\ref{ab_relation}) still holds true in the Kerr black holes with nonspherically symmetric metric. This relation  can be very useful to obtain the formulas of  the travel time of the light, which in turn generate the relativistic images due to the black holes in a more geometric way represented by the impact parameter $b_{sc}$ of the light sphere. It will be of great interest to follow \cite{Gralla_2020b} obtaining the coefficients $\tilde a$, $\tilde b$ and $\bar a$, $\bar b$ on the equatorial plane and then extending our studies on the general nonequatorial plane.
}

Although  (\ref{tilde_b_k}) still looks complicated, our approach provides a systematical way to compute not only $\tilde a$ and $\bar a$ but also $\tilde b$ and $\bar b$.
It is clear that the coefficient $\tilde b$ in  (\ref{tilde_b_k}) has strong dependence of the distance of the observer and the source measured from the black hole, where $Z_{obs}=1-r_0/R_{obs} \to 1$ and $Z_{src}=1-r_0/R_{src} \to 1$ for $R_{obs},R_{src} \gg r_0$.
In particular, $\tilde b$ is mainly dominated by $1/(1-Z)$ and $\log (1-Z)$ evaluated at the locations of the source and the observer, respectively.
Those main contributions can be understood from the integrand (\ref{dt/dz_k}) due to  the dependence of $1/(1-z)^2$ and $1/(1-z)$. In the observation of the time delay between relativistic images, the difference in $\tilde b$ given by
two distinct trajectories of the light rays, i.e. $\Delta \tilde b$ becomes relevant. In the case of  the Schwarzschild black holes by sending $a\to 0$ in (\ref{tilde_b_k}), since $\tilde b$  just depends on the distance of the observer and the source from the black hole, $\Delta \tilde b=0$. However, as for the Kerr black holes,  the contributions of  the large $1/(1-Z)$ terms in $\tilde b$ from the distinct trajectories of the light rays  are canceled  so that $\Delta \tilde b$  depends only on the black hole's parameters.

In what follows, two interesting limits on the spin of the black hole $a$, namely $  a\to 0$ and $a\to M$  are considered to obtain the approximate expressions of $\tilde a$ and $\tilde b$.
In the small $a$ limit,  to its first order,
$r_{sc} \simeq  3M -2\sqrt{3}M/3\frac{}{} \left(\frac{sa}{M} \right) +\mathcal{O}\left(\frac{a}{M} \right)^2$,
$b_{sc} \simeq  3\sqrt{3}M -2M\left(\frac{sa}{M}\right) +\mathcal{O}\left(\frac{a}{M}\right)^2$, $C_{-sc} \simeq  \mathcal{O}\left(\frac{ a}{M}\right)^3$,
$C_{+sc} \simeq  2/3 +2\sqrt{3}/27 \left(\frac{s a}{M}\right) +\mathcal{O}\left(\frac{a}{M}\right)^3$, and $\tilde{C}_{-sc}/M^2 \simeq  \mathcal{O}\left(\frac{a}{M}\right)^3$,
$\tilde{C}_{+sc}/M^2 \simeq  2\sqrt{3} \left(\frac{s a}{M}\right) + \mathcal{O}\left(\frac{a}{M}\right)^2$, giving
\begin{eqnarray}
\bar{a} & \simeq & 1 +\frac{2\sqrt{3}}{9} \left(\frac{s a}{M}\right) +\mathcal{O}\left(\frac{a}{M}\right)^2\; , \label{bar_a_a_k} \\
\bar{b} & \simeq & -\pi +\log{[ 216(7-4\sqrt{3}) ]} \nonumber \\
  &&+ \frac{2}{3\sqrt{3}}\left(1+\log{[216(7-4\sqrt{3})]} +\frac{3\sqrt{3}}{2}\log{\frac{216(7-4\sqrt{3})}{72}} \right) \left(\frac{s a}{M}\right)
 +\mathcal{O}\left(\frac{a}{M}\right)^2 ,\label{bar_b_a_k}\\
\frac{\tilde{a}}{M} & \simeq & \frac{3\sqrt{3}}{2} +\frac{17\sqrt{3}}{18}\left(\frac{a}{M}\right)^2 +\mathcal{O}\left(\frac{a}{M}\right)^3 \; ,\label{tilde_a_a_k}\\
\frac{\tilde{b}}{M} & \simeq & \frac{R}{M} + 2 \log{\left[(26-15\sqrt{3})\left(3+2\frac{R}{M}\right)\right]}
-3\sqrt{3} \left[ 1+\log{\left( \frac{2\sqrt{6}+3\sqrt{2}}{36} \right)}  \right] \nonumber\\ &&+ \left(\frac{sa}{M}\right) +\mathcal{O}\left(\frac{ a}{M}\right)^2 \; .
 \label{tilde_b_a_k}
\end{eqnarray}
 Since the time delay between two relativistic images depends on the difference in $\tilde b$, for both light rays traveling along either direct or retrograde orbits, $\Delta \tilde b=0$ and for one light ray along
the direct orbit and the other along the retrograde orbit  $\Delta \tilde b=\tilde{b}(a)-\tilde{b}(-a)$ becomes
 \begin{equation}\label{delta_tilde_b_a}
\frac{\Delta \tilde b}{M}=\frac{\tilde{b}(a)}{M}-\frac{\tilde{b}(-a)}{M}
\simeq  2 \left(\frac{a}{M}\right)+ \mathcal{O}\left(\frac{ a}{M}\right)^2 \, .
\end{equation}
To the order of $a/M$, the dependence of $R$ is removed, and $\Delta \tilde b$ has the linear dependence in $a$, solely depending on the black hole's parameters.

As $a\to M$  in the extreme black hole, the corresponding event horizon collapse to $r_{\pm}=M$ and the radius of the light sphere turns out to be the same as the event horizon $r_{+c}=M$.
The extremely strong gravitational field in the direct orbits will dramatically affect the time delay observations as compared with  the retrograde orbits, where the radius of the light sphere is large than the event horizon, $r_{-c}=4M > r_{\pm}$.
It deserves more efforts to find the approximate expressions of $\tilde a$, $\tilde b$ and $\bar a$, $\bar b$
in the extreme black hole  by introducing a small parameter $\epsilon =1-a/M$.
For $\epsilon \to 0$,
$r_{sc} r_{\pm}-a^2 \simeq \mathcal{O}(\sqrt{\epsilon})$,
$C_{-sc} \simeq 1/2-\sqrt{3}/4 +\mathcal{O}(\sqrt{\epsilon})$,
$C_{+sc} \simeq  1/2 + \sqrt{3}/{4} +\mathcal{O}(\sqrt{\epsilon})$,
$\tilde{C}_{-sc}/M^2 \simeq  -\sqrt{2}/4/\sqrt{\epsilon}  +(1-2/\sqrt{3}) +\mathcal{O}(\sqrt{\epsilon})$ ,
and
$ \tilde{C}_{+sc}/M^2 \simeq  \sqrt{2}/4/\sqrt{\epsilon}  +(1+2/\sqrt{3}) +\mathcal{O}(\sqrt{\epsilon})$
for $s=+1$ in the direct orbits.
As long as $\sqrt{\epsilon} < 1-Z =M/R \ll 1$, where $R$ denotes the distance of either the source or the observer measured from the black hole, the behavior of the coefficients $\tilde a$, $\tilde b$, $\bar a$ and $\bar b$ near the extreme black holes is found to be
\begin{eqnarray}
\bar{a} & \simeq & \frac{1}{\sqrt{2}\sqrt{\epsilon}} +\frac{11\sqrt{3}}{6} +\mathcal{O}(\sqrt{\epsilon}) \;, \label{bar_a_+_epsilon_k} \\
\bar{b} & \simeq & \frac{1}{\sqrt{2}\sqrt{\epsilon}} \left[ \log{\left( \frac{3\sqrt{6} }{8} \sqrt{\epsilon}\right)}-\frac{6~192~495~697~115~372}{5~636~787~964~422~615} \right]  +\frac{25\sqrt{3}}{36} \log{(\epsilon)}+ \mathcal{O}({\rm const.})\;,  \label{bar_b_k} \label{bar_b_+_epsilon_k}\\
\frac{\tilde{a}}{M} & \simeq & \frac{1}{\sqrt{2} \sqrt{\epsilon}} +7\sqrt{3}  +\mathcal{O}(\sqrt{\epsilon})\; , \label{tilde_a_+_epsilon_k}\\
\frac{\tilde{b}}{M} & \simeq & \frac{1}{\sqrt{2}\sqrt{\epsilon}} \left[  \log{\left( \frac{3\sqrt{6}}{8} \sqrt{\epsilon} \right)}-\frac{6~192~495~697~115~372}{5~636~787~964~422~615}   \right]+ \frac{55}{48\sqrt{3}} \log{(\epsilon)} \nonumber \\
&& +  \frac{R}{M} + 2 \log{\left(\frac{R}{M}\right)} + \mathcal{O}({\rm const.})\; .\label{tilde_b_+_epsilon_k}
\end{eqnarray}
Again, since $r_{\pm}=r_{+c}=M$ for direct orbits when $ a\to M$,  the  strong gravitational effects on the light rays drive  the values of $\bar a $, $\vert \bar b \vert $, $\tilde a$ and $\vert \tilde b \vert $  to infinity in the way that the ratio $\tilde a/\bar a$ remains finite.

For the case of $s=-1$ along the retrograde orbits, the corresponding $r_{-c}=4 M $ when $a\to M$ is larger than the event horizon $r_{\pm}=M$.
The relevant coefficients are approximately by
$C_{-sc} \simeq  -9\sqrt{2}/112/\sqrt{\epsilon} +2/7 +\mathcal{O}(\sqrt{\epsilon})$,
$C_{+sc} \simeq  9\sqrt{2}/112/\sqrt{\epsilon} +2/7 +\mathcal{O}(\sqrt{\epsilon}),$ and
$\tilde{C}_{-sc}/M^2 \simeq  17\sqrt{2}/16/\sqrt{\epsilon} -2 +\mathcal{O}(\sqrt{\epsilon})$,
$\tilde{C}_{+sc}/M^2 \simeq - 17\sqrt{2}/16/\sqrt{\epsilon} -2 +\mathcal{O}(\sqrt{\epsilon})$.
Near the extreme black hole,
\begin{eqnarray}
\bar{a} &\simeq & \frac{4}{3\sqrt{3}} +\mathcal{O}(\epsilon) \; ,\label{bar_a_-_epsilon_k}\\
\bar{b} &\simeq & -\pi + \frac{1}{9} \left[ -6+8\sqrt{3} +4\sqrt{3} \log{\left(\frac{1152}{7}(7-4\sqrt{3})\right)} \right] +\mathcal{O}(\epsilon) \; , \label{bar_b_-_epsilon_k}\\
\frac{\tilde{a}}{M} &\simeq & \frac{14 }{3\sqrt{3}} +\mathcal{O}(\epsilon) \; ,\label{tilde_a_-_epsilon_k}\\
\frac{\tilde{b}}{M} &\simeq & \frac{R}{M} +2 \log{\left( \frac{2-\sqrt{3}}{2} \frac{R}{M} \right)} + \mathcal{O}(\sqrt{\epsilon})\; .\label{tilde_b_-_epsilon_k}
\end{eqnarray}
All of them are finite as expected.

\begin{figure}[h]
\centering
\includegraphics[width=0.88\columnwidth=0.88,trim=200 0 200 0,clip]{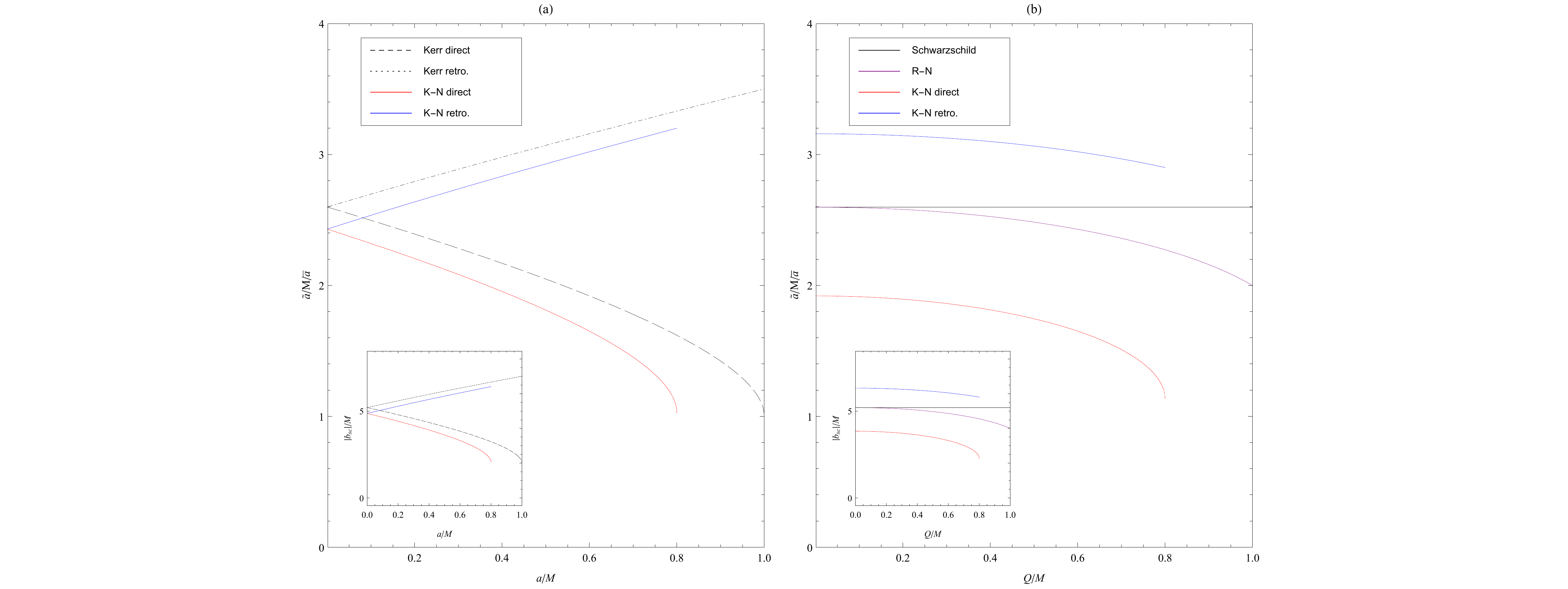}
 \caption{
{The ratio  $\tilde{a}/M/\bar{a}$ as a function of the spin parameter $a/M$ and the black hole charge $Q/M$.
%
%
The left and right panels show the cases of Kerr-Newman black holes with $Q/M=0.6$ and $a/M=0.6$, respectively.
The cases for Kerr, Reissner-Nordstr\"om, and Schwarzschild black holes are shown for comparison.
The insets show $ b_{sc} $ versus $a/M$ and $Q/M$ for comparison. } }
\label{fig2}
\end{figure}

\begin{figure}[h]
\centering
\includegraphics[width=0.88\columnwidth=0.88,trim=200 0 200 0,clip]{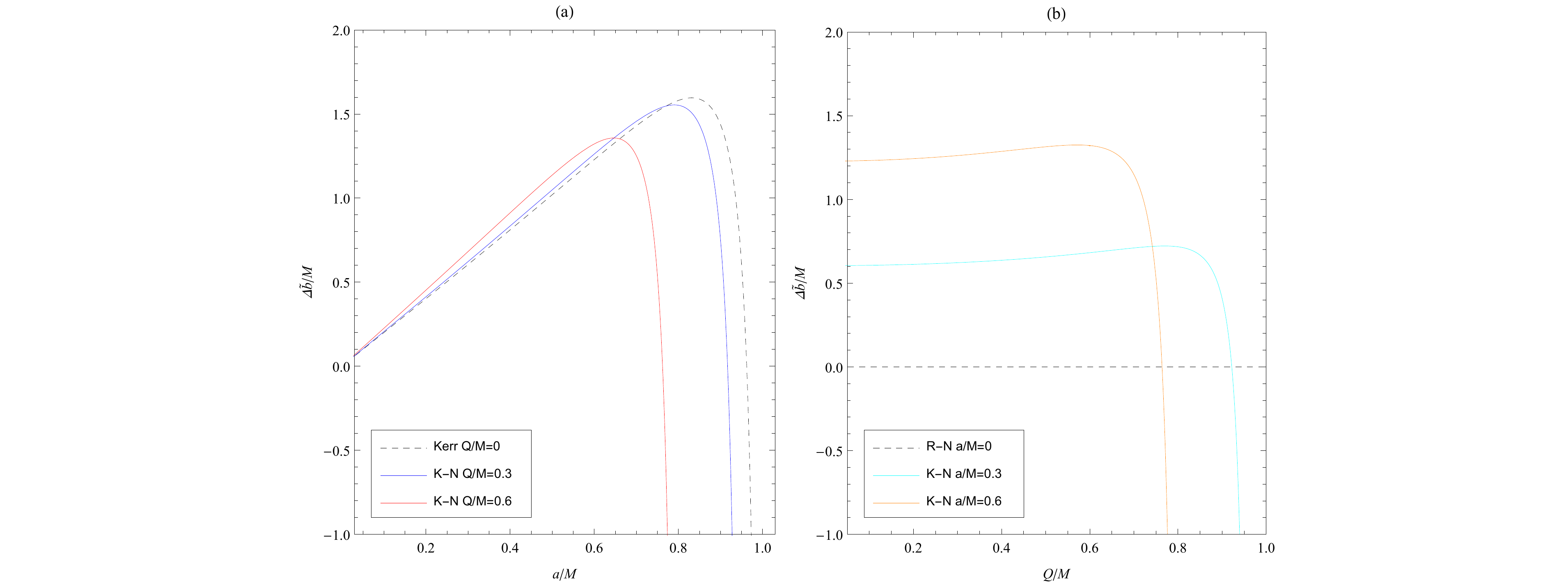}
 \caption{
{The difference $\tilde{b}(a)-\tilde{b}(-a)$ versus the spin parameter $a/M$ and the black hole charge $Q/M$.
Plot (a) shows the cases $Q/M=0.3$ (blue line) and $Q/M=0.6$ (red line) for Kerr-Newman black holes. The result of the Kerr black hole (dashed line) is also shown for comparison.
Plot (b) shows the results of $a/M=0.3$ (cyan line) and $a=0.6$ (orange line) for Kerr-Newman black holes. The result of the Reissner-Nordstr\"om (dashed line) is shown for comparison.}
}
\label{fig3}
\end{figure}

We will show later that the time delay from two distinct orbits circling around the extreme black hole will depend on
 $\tilde a/\bar a$,  $\bar b $ and $\Delta \tilde b$.
Thus, in addition to $\tilde a$ and $\tilde b$,    $\bar a$ and $\bar b$ given by (\ref{abar_k}) and  (\ref{bbar_k})  are also very relevant to the time delay calculations. Notice that with the parameters under investigation $\bar a>0$, but $\bar b<0$.
Our results are shown in \cite{Hsiao_2020}, where  both $\bar a$ and $\vert \bar b \vert$ increase (decrease) in $a$ in  direct (retrograde) orbits, giving the fact that the deflection angle $\hat \alpha$ decreases (increases) with the increase of the black hole's spin for a given impact parameter.
%
{
The behavior of $\bar b$ as a function of $a$ has been  drown in  \cite{Hsiao_2020}, in which  as $a\to M$ in the extreme black hole cases  for the direct orbit, $\vert \bar b\vert \propto (1/\sqrt{\epsilon}) \log[1/\sqrt{\epsilon}]$ as $\epsilon \to 0$ whereas for the retrograde orbit $\bar b $ remains a finite value, consistent with our analytical analysis above in (\ref{bar_b_+_epsilon_k}) and (\ref{bar_b_-_epsilon_k}).}
Fig.(\ref{fig2}) is plotted for the ratio of $\tilde a/\bar a$ that decreases with $a$ for direct orbits but increases in stead for retrograde orbits.
Both of them reach a finite value as $a\to M$.
%
In fact, in the Kerr case,  given the effective potential  $W_\text{eff}$ in (\ref{eq:Weff_k}) we have stated  that the nonzero spin of the black holes induces more repulsive  effects  to the light rays in the direct  orbits than in the retrograde orbits so as to decrease  $b_{+c}$  but to increase  $b_{-c}$ as $a$ increases.
Also,  $\Delta \tilde b=\tilde b(+a)-\tilde b(-a)$ between the direct orbit and the retrograde orbit is plotted in
Fig.(\ref{fig3}) where for small $a$, $\Delta \tilde b$ increases with $a$ consistent with (\ref{delta_tilde_b_a}), and then goes to $- \infty$ mainly resulting from strong gravitational time delay effects in the direct orbits with $\tilde b(+a) \propto (1/\sqrt{\epsilon}) \log[\sqrt{\epsilon}] $ for small $\epsilon$ in (\ref{tilde_b_+_epsilon_k}).


\subsection{Kerr-Newman black holes}

The line element given by the Kerr-Newman metric of the nonspherically symmetric spacetime with angular momentum $a$ and charge $Q$   that we would also like to explore is
\ba
 {ds}^2 &=& g_{\mu\nu} dx^\mu dx^\nu  \nonumber \\
&= & -\frac{ \left(\Delta -a^2 \sin^2\theta \right)}{\Sigma } {dt}^2   + \frac{ a   \sin ^2\theta \left(Q^2-2 M
   r\right)}{\Sigma } ({dt}{d\phi+ d\phi dt)}  \nonumber\\
&&+  \frac{\Sigma}{\Delta} dr^2 +\Sigma{\, d\theta}^2 +\frac{ \sin ^2 \theta}{\Sigma} \left(( r^2+a^2)^2 -a^2 \Delta  \sin
   ^2\theta \right) {d\phi}^2 \, , \label{KN_metric}
   \ea
where
\be \label{Delta_kn}
\Sigma=r^2+a^2\cos^2\theta \, , \quad  \Delta=r^2+a^2+Q^2-2 M r\,.
\ee
In the Kerr-Newmann cases, the corresponding outer (inner) event horizon $r_+$ ($r_-$) is
\be \label{rpm_kn}
r_{\pm}=M\pm\sqrt{M^2-(Q^2+a^2)}\,
\ee
with  $M^2 > Q^2+ a^2$.

The effective potential  function $W_\text{eff}$, the counterpart of  (\ref{eq:Weff_k}),  is obtained as
 \begin{align}
W_\text{eff}(r)=\frac{1}{r^2}
\left[1-\frac{a^2}{b^2}+\left(-\frac{2M}{r}+\frac{Q^2}{r^2}\right)\left(1-\frac{a}{b_s}\right)^2\right] \, . \label{eq:Weff_kn}
\end{align}
The nonzero charge of the black hole  give repulsive effects  to light rays as seen from  the $1/r^4$ term, which shifts  the innermost circular trajectories  of the light  toward the black holes with the smaller critical impact parameter $b_{sc}$ for both direct and retrograde orbits, leading to the smaller radius of the light sphere $r_{sc}$.
The travel time of the light  crucially depends not only on the locations of the source and the observer but also on the radius of the light sphere circling around  the black hole.

The equation of motion for the light rays in terms of the variable $z$ similar to (\ref{dt/dz_k}) in the Kerr case  is obtained as
\begin{equation}
\frac{dt}{dz}=\frac{r_0}{(1-z)^2} \left[ \frac{a (1-\frac{a}{b}) (1-z)^3 (-\frac{2M}{r_0}+\frac{Q^2}{r_0^2}(1-z))+ \frac{r_0^2}{b} + \frac{a^2}{b}(1-z)^2}{1-\frac{2M}{r_0}(1-z)+\frac{a^2+Q^2}{r_0^2}(1-z)^2}  \right]\frac{1}{\sqrt{B(z,r_0)}}\;,
\end{equation}
where
\begin{equation}
B(z,r_0)=c_1(r_0)z+c_2(r_0)z^2+c_3(r_0)z^3+c_4(r_0)z^4 \, .
\end{equation}
The function $B(z,r_0)$ is a quartic polynomial in $z$ with the coefficients
\begin{eqnarray}
c_1(r_0)&=&4Q^2\left(1-\frac{a}{b_s}\right)^2-6Mr_0\left(1-\frac{a}{b_s}\right)^2+2r_0^2\left(1-\frac{a^2}{b^2}\right)\,, \label{c1_kn}\\
c_2(r_0)&=&-6Q^2\left(1-\frac{a}{b_s}\right)^2+6Mr_0\left(1-\frac{a}{b_s}\right)^2-r_0^2\left(1-\frac{a^2}{b^2}\right)\, , \label{c2_kn}\\
c_3(r_0)&=&4Q^2\left(1-\frac{a}{b_s}\right)^2-2Mr_0\left(1-\frac{a}{b_s}\right)^2\,,\label{c3_kn} \\
c_4(r_0)&=&-Q^2\left(1-\frac{a}{b_s}\right)^2  \,  .  \label{c4_kn}
\end{eqnarray}
All coefficients have the additional dependence of the black hole charge $Q$. The presence of the $z^4$ term with the coefficient $c_4(r_0)$, which vanishes in  the Kerr case, make the expressions of $\tilde a$ and $\tilde b$ more lengthy.

The integrand $f(z,r_0)$ in (\ref{travel_time})  now takes the form
\begin{equation}\label{f_kn}
\begin{split}
f(z,r_0)=& \frac{r_0}{b} \frac{r_0^2}{a^2+Q^2} \left[ \frac{\tilde C_-}{z-z_-}+\frac{\tilde C_Q z+\tilde C_+}{z-z_+}+ \frac{r_0^2}{(1-z)^2 (z-z_-) (z-z_+)} \right] \\
&\quad\quad\quad
\times\frac{1}{\sqrt{c_1(r_0)z+c_2(r_0)z^2+c_3(r_0)z^3+c_4(r_0)z^4}} \,.
\end{split}
\end{equation}
The corresponding coefficients $\tilde C_-$, $\tilde C_Q$, $\tilde C_+$ in the Kerr-Newman case are
\begin{eqnarray} \label{C_tilde_pmQ_kn}
\tilde C_-&=&{\frac{-2aMbr_- +a^2(a^2+Q^2) +2a^2Mr_- +a(b-a)\frac{Q^2}{r_0^2}\frac{r_0^2r_-^2}{a^2+Q^2}}{2r_0\sqrt{M^2-a^2-Q^2}}}\, ,\\
\tilde C_Q&=& \frac{a Q^2}{r_0^2}\left(1-\frac{a}{b_s}\right) b\, ,\\
\tilde C_+&=&{\frac{2aMbr_+ -a^2(a^2+Q^2) -2a^2Mr_+ +a (a-b)\frac{Q^2}{r_0^2}(-r_0r_-+r_0r_++\frac{r_0^2}{a^2+Q^2}r_-r_+)}{2r_0\sqrt{M^2-a^2-Q^2}}} \, ,
\end{eqnarray}
Also  $z_+$, $z_-$ now become
\begin{eqnarray}
z_-&=&1-\frac{r_0r_-}{a^2+Q^2}\;,\\
z_+&=&1-\frac{r_0r_+}{a^2+Q^2}\;,
\end{eqnarray}
defined in terms of  the outer(inner) black hole horizon $r_+$ ($r_-$). Again, $z_\pm \le 0$ for all $a$ and $Q$ with the nonzero $r_+$. Note that, for charge $Q \to 0$, $\tilde C_Q$ vanishes.

Following the previous study of the Kerr black holes, we define the function $f_D(z,r_0)$ as
\begin{equation}
f_D(z,r_0)=\frac{r_0^3}{b(a^2+Q^2)}\left[ \frac{\tilde C_-}{z-z_-}+\frac{\tilde C_Q z+\tilde C_+}{z-z_+} + \frac{r_0^2}{(1-z)^2 (z-z_-) (z-z_+)}  \right] \frac{1}{\sqrt{c_1 z+c_2 z^2}}\, ,
\end{equation}
and its integration over $z$ gives $\tilde a$ and $\tilde b_D$ in this case. Also, $f_R(z,r_0)=f(z,r_0)-f_D(z,r_0)$ and  the corresponding integral of $f_R$ leads to $\tilde b_R$.
Finally, the coefficients $\tilde a$ and $\tilde b$  in the Kerr-Newman case become
\begin{equation}\label{tilde_a_kn_f}
\begin{split}
\tilde{a}=& \frac{r_{sc}^3}{2 b_{sc} \sqrt{c_{2sc}}}
\left[ \frac{2M+r_{sc}}{2r_{sc}}+ \frac{\tilde C_{- sc}}{r_{sc}r_{-}-a^2-Q^2}
+\frac{(a^2+Q^2)^3}{r_-^2 (r_{sc}r_- -a^2-Q^2) r_{sc} (r_+-r_-)}
 \right]+(- \leftrightarrow +) \, ,
\end{split}
\end{equation}
{where by replacing $b_{sc}$ with (\ref{bsc_kn2}), and using (\ref{light_sphere_kn}), we can simplify the above expression to be
\begin{equation}\label{tilde_a_kn}
\tilde a = \frac{1}{2} \frac{\sqrt{M r_{sc}-Q^2}}{\sqrt{3 M r_{sc}- 4 Q^2}} \left[ \frac{ r_{sc}(r_{sc}+3M)}{r_{sc}-M} -2Q^2 \right]  \,.
\end{equation}}
and
\begin{equation}\label{tilde_b_kn}
\begin{split}
&\tilde{b}=\frac{1}{2}\tilde{a} \log{\left[ \frac{8c_{2sc}^2 Z^2  b_{sc}''}{c_{1sc}'^2 b_{sc}} \right]}\\
&+\frac{r_{0}^3}{2 b \sqrt{c_{2}} }  \left[ \frac{\sqrt{3}\sqrt{(3-2Z)+(-4Z+3Z^2)(c_{4}/c_{2})} }{(1-c_{4}/c_{2})(1-Z)}\right.\\
&\left.
 + \frac{2 \sqrt{3}M}{r_{0} \sqrt{1-c_{4}/c_{2}}} \log{\left(  \frac{(2-Z)(1-c_{4}/c_{2})+\sqrt{1-c_{4}/c_{2}}\sqrt{(3-2Z)+(-4Z+3Z^2)(c_{4}/c_{2})} }{2(1-c_{4}/c_{2})+\sqrt{3}\sqrt{1-c_{4}/c_{2}}} \frac{1}{1-Z} \right)}\right.\\
&\left.\quad\quad\quad\quad -\frac{3}{(1-c_{4}/c_{2})} - \frac{r_{0}^2}{(a^2+Q^2-2Mr_{0}+r_{0}^2)} \log \tilde{N}(Z,a,Q) \right] \\
&+\frac{r_{0}^3}{b\sqrt{c_{2}} } \frac{\tilde C_{-}}{a^2+Q^2-r_{0}r_-}\left[ \log\tilde{N}(Z,a,Q) -\frac{\sqrt{3}(a^2+Q^2)}{\tilde{W}_-}\log\tilde{M}(Z,a,Q,r_{-})\right] \\
&-\frac{r_{0}^2}{b\sqrt{c_{2}} }  \frac{\sqrt{3}(a^2+Q^2)^4}{(a^2+Q^2-r_{0}r_-)r_-^2(r_+-r_-)\tilde{W}_-}
\log\tilde{M}(Z,a,Q,r_{-}) +(- \leftrightarrow +)\\
&+\frac{r_{0}^3}{b\sqrt{c_{2}} }\frac{\sqrt{3}\tilde C_{Q }}{\tilde{W}_+} \log\tilde{M}(Z,a,Q,r_{+}) \vert_{r_0=r_{sc}} \;,
\end{split}
\end{equation}
where
\begin{eqnarray}
&& \tilde{M}(Z,a,Q,r_{\pm}) = \frac{a^2+Q^2-r_{0}r_{\pm}}{(a^2+Q^2)(1-Z)-r_{0}r_{\pm}}  \frac{\tilde{A}}{\tilde{B}} \; , \\
&&\tilde{A}= (2-Z)(a^2+Q^2)+r_{0}r_{\pm} +[(-2+Z)(a^2+Q^2)+(2-3Z)r_{0}r_{\pm}]\frac{c_{4}}{c_{2}} \nonumber\\
&&\quad\quad\quad\quad\quad\quad\quad\quad\quad+\sqrt{(3-2Z)+(-4Z+3Z^2)\frac{c_{4}}{c_{2}}} \tilde{W}_{\pm} \; ,\\
&&\tilde{B}= 2(a^2+Q^2)+r_{0}r_{\pm} - 2(a^2+Q^2+r_{0}r_{\pm}]\frac{c_{4}}{c_{2}}+\sqrt{3}\tilde{W}_{\pm}
\end{eqnarray}
with
\begin{equation}
\tilde{W}^2_{\pm}=(a^2+Q^2+2r_{0}r_{\pm})(a^2+Q^2)-(a^2+Q^2+3r_{0}r_{\pm})(a^2+Q^2-r_{0}r_{\pm})\frac{c_{4}}{c_{2}} \;
\end{equation}
and
\begin{equation}
\tilde{N}(Z,a,Q)=\frac{(3-Z)-2Z(c_{4}/c_{2})+\sqrt{3} \sqrt{(3-2Z)+(-4Z+3Z^2) (c_{4}/c_{2})}}{6}\;.
\end{equation}
In the equations above, we have replaced $c_{3sc} $ by the linear combination of $c_{2sc}$ and $c_{4sc}$
with
\begin{equation}
\begin{split}
c_{3sc}=&-\frac{2}{3}c_{2sc}-\frac{4}{3}c_{4sc} \, .
\end{split}
\end{equation}
{Similar to the way to simplify the coefficient $\tilde a$, the expression of the coefficient $\bar a$ in (\ref{abar_kn_f}) can be more compact to be
\begin{equation}\label{abar_kn}
\bar a = \frac{2(Q^2 -M r_{sc})}{(M-r_{sc}) \sqrt{3 M r_{sc}-4Q^2}}  \,.
\end{equation}}

Now we consider the limit of $a\to 0$ for the Reissner-Nordstr$\ddot{\rm o}$m black hole with the spherical symmetry metric.
As $ a\to 0$,  using the known equation for the radius of the light sphere $r_{c}^2=-2Q^2+3Mr_{c}$ \cite{Tsukamoto_2017a} gives
the coefficient $\bar a$ in (\ref{abar_kn}) for the Reissner-Nordstr$\ddot{\rm o}$m case becoming
\begin{equation}\label{bar_a_rn}
\bar{a}=\frac{r_{c}}{\sqrt{3M r_{c}-4 Q^2}}\;,
\end{equation}
where  the subscript is changed from $sc$ to $c$ since the same radius of the light sphere is obtained for light rays in direct orbits and retrograde orbits in the case of the non-spinning black holes.
The expected relation in (\ref{ab_relation}) can be achieved from (\ref{tilde_a_kn}) and (\ref{bar_a_rn})
with the formulas of the critical radius
\begin{equation}
r_{c}=\frac{3M+\sqrt{9M^2-8Q^2}}{2}\;,
\end{equation}
 for the spherically symmetric  black hole.
It is quite straightforward to find that the relation  (\ref{ab_relation}) is  also fulfilled from (\ref{tilde_a_kn}) and (\ref{abar_kn})
in the Kerr-Newman black holes.  This is one of the interesting findings in this work.   The formula of the time delay to be developed later can be determined by the impact parameter of the light sphere $b_{sc}$.

As $Q\to 0$, $\tilde C_Q$ and $ c_4 \to 0$. The coefficients $\tilde a $ and $\tilde b$ in (\ref{tilde_a_kn}) and (\ref{tilde_b_kn}) of the Kerr-Newman black holes reduce to that of the Kerr black holes in (\ref{tilde_a_k}) and (\ref{tilde_b_k}), respectively.
One of the applications  of the analytical expressions of $\tilde a$ and $\tilde b$
is to find their values with the input of $b_{sc}$ in (\ref{bsc_kn}) or $r_{sc}$ in (\ref{rsc_kn}) for given black hole's parameters.
As in the Kerr case, it is quite interesting to study  the limiting cases  of either $a\to 0$ or $a\to \sqrt{M^2-Q^2}$, with which to analytically understand the time delay effects.
We find that, in  both limits,  $\tilde a$, $\tilde b$, $\bar a$ and $\bar b$ follow the same expansion
behavior as in (\ref{bar_a_a_k})-(\ref{tilde_b_a_k}) when $a \to 0$ and (\ref{bar_a_+_epsilon_k})-(\ref{tilde_b_-_epsilon_k})  when $a\to \sqrt{M^2-Q^2}$ with the additional dependence on the charge $Q$ in the expansion coefficients.
Nevertheless,  the expressions of those coefficients are too lengthy to show them here.
%
{With that features in both limits allows us to see  the charge $Q$ effects in Figs (\ref{fig2}) and (\ref{fig3}).}
Due to the fact that the bending angle for light rays resulting from the charged black hole is suppressed as compared with the neutral   black hole with the same impact parameter $b$ and the angular momentum of the black holes $a$, both $\bar a $ in (\ref{abar_kn}) and $\vert \bar b \vert $ in (\ref{bbar_kn}) are found to increase with the charge $Q$ shown in our previous paper \cite{Hsiao_2020}.
%
With fixed $Q$, for the direct orbit $\vert \bar b\vert $ increases with $a$ and then reaches infinity as the spin reaches its extreme value, $a\to \sqrt{M^2-Q^2}$ and for the retrograde orbit, $\vert \bar b\vert $ decreases with $a$ all the way to a finite value in the extreme black holes as in the Kerr cases in (\ref{bar_b_+_epsilon_k}) and (\ref{bar_b_-_epsilon_k}), which have been shown in \cite{Hsiao_2020}.
In Fig.(\ref{fig2}), given $Q$  the ratio  $\tilde a/\bar a$ behaves the same as in the Kerr cases except that the largest value of $a$ can be $a=\sqrt{M^2-Q^2}$ in the extreme black holes.
Since the nonzero charge of the black hole  give repulsive effects  to light rays giving that $ b_{sc}$ decreases with $Q$ for both direct and retrograde orbits, the parameter $\tilde a/\bar a $ due to the relation (\ref{ab_relation})  also decreases with $Q$.
Fig.(\ref{fig3}) shows  $\Delta \tilde b$ as a function of $Q$ ($a$) with a fixed $a$ ($Q$)  in that $\Delta \tilde b$ respectively goes to $- \infty$  in the extreme Kerr-Newman  black holes  ($\Delta \tilde b \propto \log\epsilon$ with $\epsilon=1-{\sqrt{ M^2+Q^2}}/{M}$ in the small $\epsilon$ limit) as in the extreme Kerr black holes.
In the next section we
will see how the charge $Q$ of the black hole affects the time delay observations.
%


\section{Time delay due to Gravitational lens of supermassive galactic black holes}

We now derive the formulas of the time delay between two relativistic images based upon (\ref{travel_time_SDL}) as well as the angular positions of those images obtained in  \cite{Hsieh_2021}.
The angular positions of the source and the image are measured from the optical axis, and are denoted by $\beta$ and $\theta$, respectively (see \cite{Hsieh_2021} for detail). The lens equation as the light rays travel around the black hole $n$ times can be simplified as
%
\begin{equation}\label{leneq_app}
s\beta\simeq \theta-\frac{d_{LS}}{d_{S}}[\hat{\alpha}(\theta)-2n\pi] \,
\end{equation}
with the relation between the impact parameter $b$ and the angular position of the image approximated by
\begin{equation}\label{b_theta}
b\simeq d_L\theta \,.
 \end{equation}
According to \cite{Hsieh_2021}, the zeroth order solution $\theta_{sn}^0$ is obtained from $\hat{\alpha}(\theta_{sn}^0)=2n\pi$. Using the SDL deflection angle in (\ref{hatalpha_as}) we have then
\begin{equation}\label{theta0}
\theta_{sn}^0=\frac{\vert b_{sc}\vert}{d_L}\left( 1+e^{\frac{\bar{b}-2n\pi}{\bar{a}}} \right)
\end{equation}
for $n=1,2,\cdots$.
{Replacing $b$ by $\theta$ via (\ref{b_theta}) in (\ref{travel_time_SDL}) and substituting the result (\ref{theta0}),
 the travel time in the SDL  becomes
\begin{equation}
\begin{split}
{T_{sn}} &= \frac{2\tilde{a}(sa)}{\bar{a}(sa)} [2\pi n  -\bar{b}(sa)] +\tilde{b}(R_{obs},sa) +\tilde{b}(R_{src},sa)\, \\
&=b_{sc} [2\pi n  -\bar{b}(sa)] +\tilde{b}(R_{obs},sa) +\tilde{b}(R_{src},sa)
\end{split}
\end{equation}
where (\ref{ab_relation}) has been used to reach the last equality.
We consider that the time delay for two relativistic images lying in the same side of the optic axis where both light rays are  along either direct or  retrograde orbits with the same $sa$ winding around the black hole $n$ and $m$ times, respectively. Their difference can be obtained as
\begin{equation}\label{delta_T_s}
\begin{split}
{\Delta T^{(s)}_{n,m} = 2\pi  b_{sc} (n-m) } \,
\end{split}
\end{equation}
with the superscript ``s'' in $T$.
In the other case, when two images are on the opposite side of the optical axis resulting from the light rays circling around the black hole $n$ time in the direct orbit and $m$ times in the retrograde orbit, respectively. The time delay denoted with the superscript ``o''  then becomes
\begin{equation}\label{delta_T_o}
\begin{split}
\Delta T^{(o)}_{n,m} = &b_{+c} [2\pi n  -\bar{b}(a)] +\tilde{b}(R_{obs},a) +\tilde{b}(R_{src},a) \\
-&b_{-c} [2\pi m -\bar{b}(-a)] -\tilde{b}(R_{obs},-a) -\tilde{b}(R_{src},-a)\, .
\end{split}
\end{equation}
%
The formula of the time delay depends on the impact parameter of the light ray for the light sphere $b_{sc}$, and  the number of the loops $n$ for the light winding around the black hole and
the coefficients $\bar b$,  that together give the effective winding angle $2\pi n -\bar b$.
The dependence also involves $\tilde b$ with strong dependence of the distances of the source and the observer from the black holes.  But the difference $\Delta\tilde b$ between two distinct trajectories of the light rays solely depends on the black hole's parameters.
%
Substituting the small $a$ behavior in the Kerr case (\ref{bar_a_a_k})-(\ref{tilde_b_a_k}), the time delays $\Delta T^{(s)}_{n,m}$ and $\Delta T^{(o)}_{n,m}$ can be approximated by
\begin{equation}
\begin{split}
\Delta T^{(s)}_{n,m} = 2\pi (n-m) M \left[ 3\sqrt{3}-2 \left( \frac{sa}{M}\right)+\mathcal{O}\left(\frac{a}{M}\right)^2\right] \, ,
\end{split}
\end{equation}
\begin{equation}\label{delta_T_o_a}
\begin{split}
\Delta T^{(o)}_{n,m} =
& 3\sqrt{3}M(n-m)\pi
+2M\left[ -2(1+n+m)\pi -3\sqrt{3} \log{\left( 3(7-4\sqrt{3}) \right)}\right] \left(\frac{a}{M}\right) \\
 & + \mathcal{O}\left(\frac{a}{M}\right)^2\, .
\end{split}
\end{equation}
In the extreme case, following small $\epsilon$ expansion in (\ref{bar_a_+_epsilon_k})-(\ref{tilde_b_-_epsilon_k})  as $a\to M$, we get
\begin{equation}
\begin{split}
\Delta T^{(s)}_{n,m}(+a) = 2\pi (n-m) M \left[ 2+\left(\frac{11\sqrt{6}}{3}-12\sqrt{6}\right)\sqrt{\epsilon}+\mathcal{O}\left(\epsilon\right)\right] \, ,
\end{split}
\end{equation}
\begin{equation}
\begin{split}
\Delta T^{(s)}_{n,m}(-a) = 2\pi (n-m) M \left(14+\mathcal{O}\left(\epsilon\right)\right)\, ,
\end{split}
\end{equation}
and
\begin{equation}\label{delta_T_o_epsilon}
\begin{split}
\Delta T^{(o)}_{n,m} = & \left[ -\frac{25\sqrt{3}M}{18}+\frac{55M}{24\sqrt{3}} \right] \log{(\epsilon)}  +\mathcal{O}({\rm const}.)\, .
\end{split}
\end{equation}

The time delay $\Delta T^{(s)}_{n,m}$ given by (\ref{delta_T_s}) is determined by $b_{sc}$ shown in Fig.(\ref{fig2}) for fixed $m$ and $n$.
The decrease (increase) of $b_{+c}$($b_{-c}$) as $ a$ increases in the direct orbits (the retrograde orbits) lead to the decrease (increase) of the time delay $\vert \Delta T^{(s)}_{n,m}(+a)\vert$  ($\vert \Delta T^{(s)}_{n,m}(-a)\vert$) as long as $n\neq m$ in agreement with the results of \cite{Bozza_2004}.
{Here  we also find that the effects of the nonzero charge of the black holes decreases $\vert \Delta T^{(s)}_{n\neq m}(\pm a )\vert$ for both direct and retrograde orbits because the impact parameter of the innermost circular motion $b_{sc}$ decreases with $Q$.}

When one light ray is along the direct orbit and the other follows retrograde orbits, the time delay $\Delta T^{(o)}_{n,m}$  is plotted in Figs.(\ref{fig4}) and (\ref{fig5}), in which number of the loops are chosen to be $n=1$ ($m=1$) for the direct (retrograde) orbits. For small $a$, $\Delta T^{(o)}_{1,1}$ is dominated by $b_{sc}$ and since $b_{+c} < b_{-c}$, $\Delta T^{(o)}_{1,1}$  is negative and  $\vert \Delta T^{(o)}_{1,1}\vert $ increase with $a$, consistent with \cite{Bozza_2004}.
Its linear $a$ dependence can be seen by the analytic approximation (\ref{delta_T_o_a}).
Nevertheless, in the extreme black holes, the large gravitational effects to the direct orbits drives $\Delta T^{(o)}_{1,1} $ becoming positive and further more goes to a very large value also consistent with (\ref{delta_T_o_epsilon}).
{The effects of the charge instead lead to the larger $\vert \Delta T^{(o)}_{1,1} \vert$ for the same $a$.}
Near extreme black holes  $\Delta T^{(o)}_{1,1} \propto \log\epsilon$, where $\epsilon=1-{\sqrt{ M^2+Q^2}}/{M}$ in the small $\epsilon$ limit. The slope of  $\Delta T^{(o)}_{1,1} \to \infty $ as $\epsilon \to 0$ for nonzero charge black holes is slightly larger than the Kerr Black holes.
Again, the time delay between two relativistic images in the limits of small $a$ as well as $a\to \sqrt{M^2-Q^2}$ can be understood from taking the appropriate limits of the obtained analytical expressions of all coefficients.

We now take the examples of a galactic black hole and the supermassive black holes such as Sagittarius A*
and M87 \cite{Ghez_1998,Schodel_2002}.  We also assume a light source at the distance with the ratio $d_{LS}/d_S = 1/2$ as shown in Tables (\ref{T_1}) and (\ref{T_2}) \cite{Richstone_1998,Fragione_2020,Tamburini_2020}.
 It is evident that the black hole with higher mass or with spin of the extreme black hole will have higher time delay.
It has also been emphasized in \cite{Bozza_2004} that the time delay measurement together with the analytical expressions of all coefficients of $\bar a$ and $\bar b$ in the SDL expression of the deflection angle (\ref{hatalpha_as}) as well as $\tilde a$ and $\tilde b$ again in the SDL of the travel time (\ref{travel_time}) can be used to estimate  the properties of the black holes including its distance from the observer.

%

\begin{figure}[h]
\centering
\includegraphics[width=0.88\columnwidth=0.88]{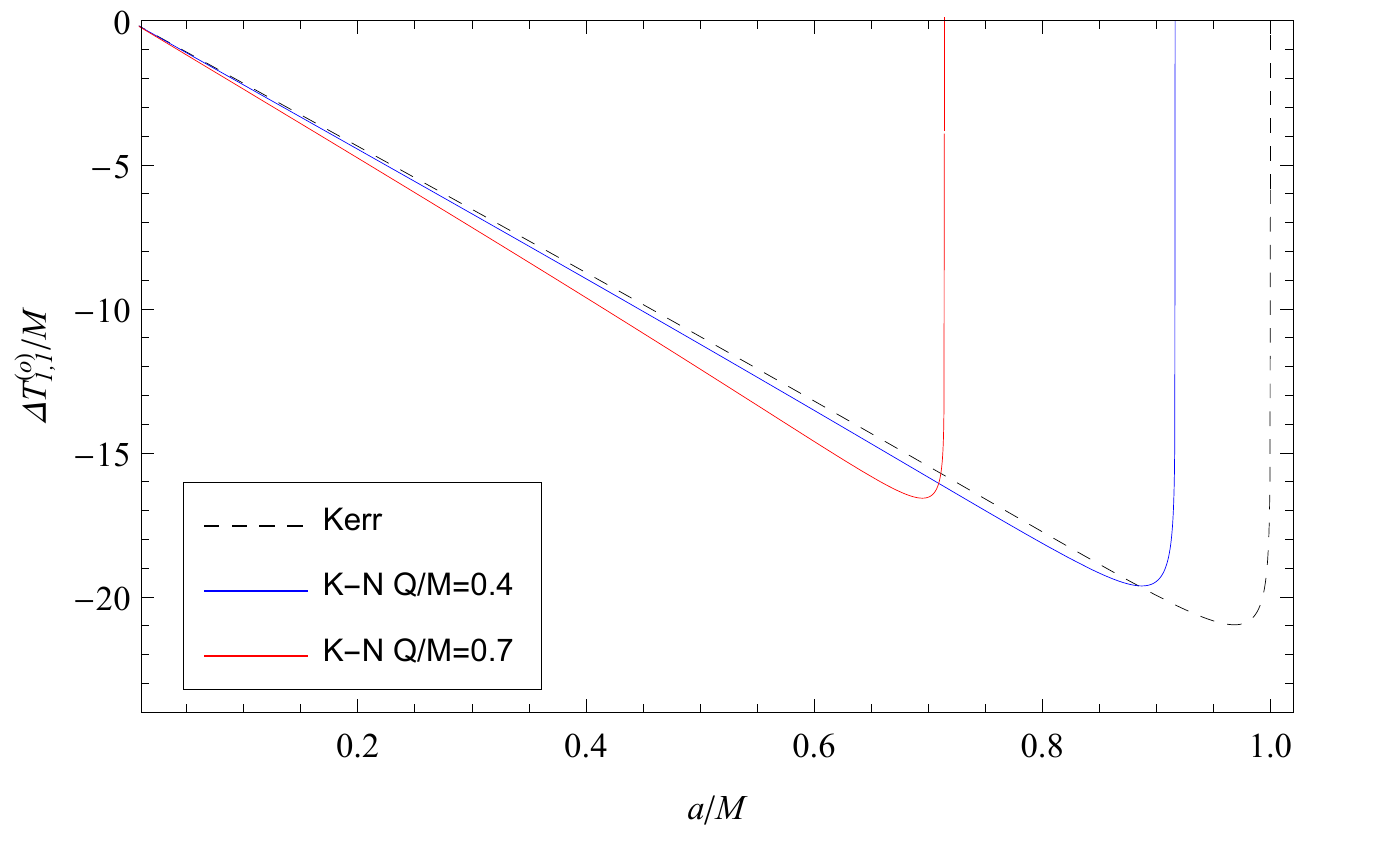}
 \caption{
{The  time delay $\Delta T^{(o)}_{{1,1}}$ as a function of the spin parameter $a/M$ with the black hole charge $Q/M=0$ (dashed line), $Q/M=0.4$ (blue line), and $Q/M=0.7$ (red line).
}
}
\label{fig4}
\end{figure}
\begin{figure}[h]
\centering
\includegraphics[width=0.88\columnwidth=0.88]{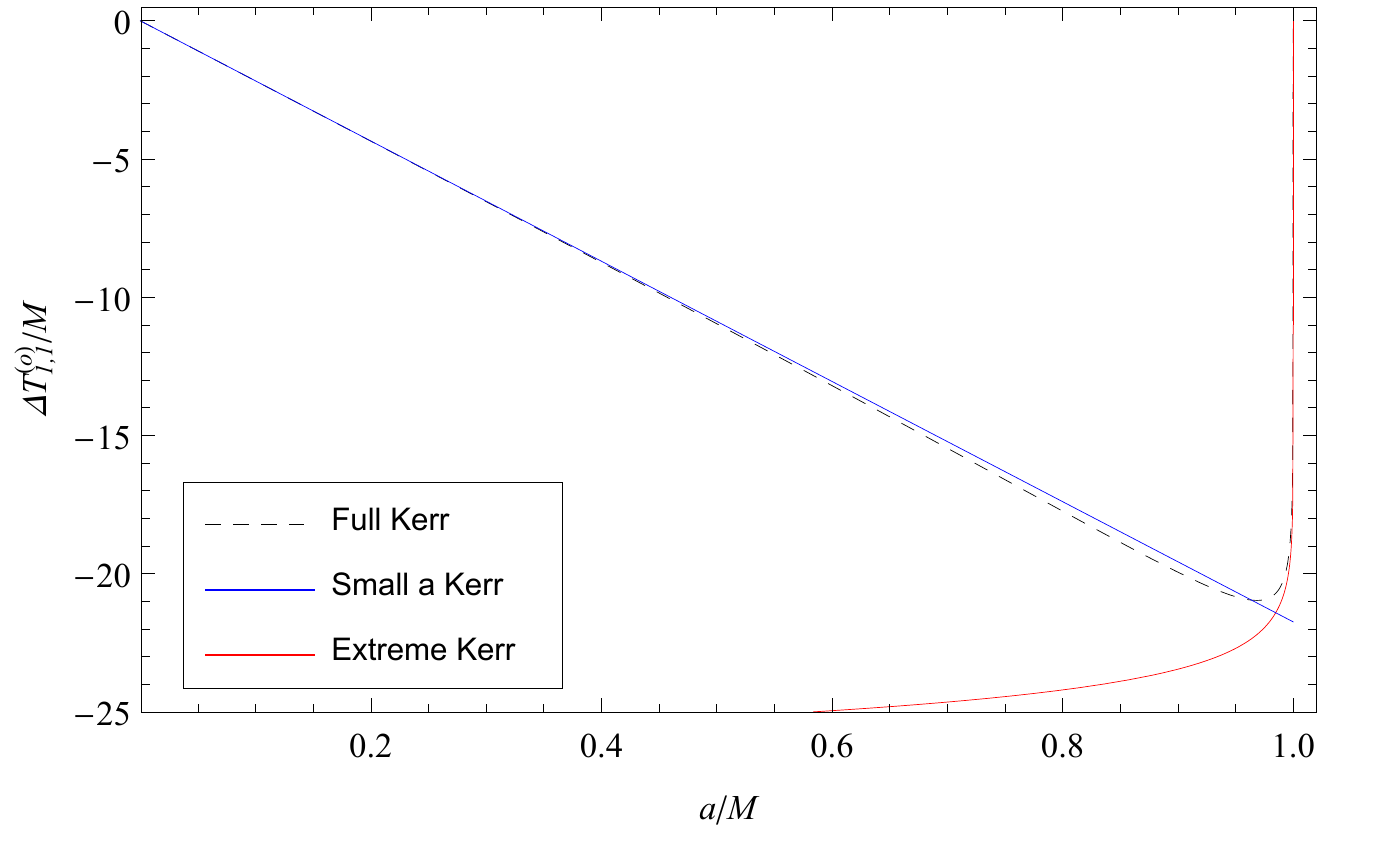}
 \caption{
{
Plot comparison of the perturbative small $a \to 0$ expansion (blue line) and the extreme Kerr approximation $a \to M $ (red line) against the full calculation of $\Delta T^{(o)}_{1,1}$.
}
}
\label{fig5}
\end{figure}

\textcolor{red}{
\begin{table}
\begin{tabular}{cccccc}
\hline
\hline
     &  $M/M_{\odot}$      & $sa/{M}$ & $Q/{M}$ & Distance $d_{L} (Mpc)$ & $\Delta T_{2,1}^{(s)} (h)$    \\
\hline
Galactic black hole    & $2.8 \times 10^6 $    & 0       & 0       & $0.0085$      & $0.125$ \\
Sagittarius A*& $4.1 \times 10^6 $    & $0.1$   & 0       & $0.0082$       & $0.176$ \\
NGC4486 (M87) & $3.3 \times 10^9 $    & $0.9$   & 0       & $15.3$        & $80.755$ \\
\hline
Galactic black hole    & $2.8 \times 10^6 $    & 0       & 0       & $0.0085$      & $0.125$ \\
Sagittarius A*& $4.1 \times 10^6 $    & $-0.1$  & 0       & $0.0082$       & $0.190$ \\
NGC4486 (M87) & $3.3 \times 10^9 $    & $-0.9$  & 0       & $15.3$        & $193.973$ \\
\hline
\hline
\end{tabular}
\caption{Estimates of the time delay $\Delta T_{2,1}^{(s)}$ for some galactic and  supermassive black holes \cite{Richstone_1998,Fragione_2020,Tamburini_2020}. The unit of the time difference $h$ means "hour". }
\label{T_1}
\end{table}
\begin{table}
\begin{tabular}{cccccc}
\hline
\hline
       &  $M/M_{\odot}$      & $a/{M}$ & $Q/{M}$ & Distance $d_{L} (Mpc)$ & $\Delta T_{1,1}^{(o)} (h)$    \\
\hline
Galactic black hole    & $2.8 \times 10^6 $    & 0       & 0       & $0.0085$      & $0$ \\
Sagittarius A*& $4.1 \times 10^6 $    & $0.1$   & 0       & $0.0082$       & $-0.012$ \\
NGC4486 (M87) & $3.3 \times 10^9 $    & $0.9$   & 0       & $15.3$        & $-90.129$ \\
\hline
\hline
\end{tabular}
\caption{Estimates of the time delay $\Delta T_{1,1}^{(o)}$ for some galactic  and supermassive black holes \cite{Richstone_1998,Fragione_2020,Tamburini_2020}. The unit of the time difference $h$ means "hour".}
\label{T_2}
\end{table}
}


\section{Summary and outlook}

In summary, we study the time delay between two relativistic images due to strong gravitational lensing caused by the Kerr and Kerr-Newman black holes respectively. Starting from the known form of the SDL deflection angle in (\ref{hatalpha_as}) with the coefficients $\bar a$ and $\bar b$, one can derive the formula for travel time of the light rays from the distant source winding around the black hole and reaching the observer, given in (\ref{travel_time}), with another two coefficients $\tilde a$ and $\tilde b$.
Here we successfully obtain their analytical expressions and together with the coefficients $\bar a$ and $\bar b$, we are able to explore the time delay phenomena.  
{The expression of the coefficient $\tilde a$ is then compared with the known results of the Schwarzschild black holes \cite{Bozza_2004}.  Also,
 we reproduce the coefficient $\tilde a$ and $\bar a$ in the Kerr black holes
 shown in \cite{Gralla_2020a,Gralla_2020b},restricting the light rays on the equatorial plane, and extend them to the Kerr-Newman black holes.  }
Based upon our analytical and numerical studies, the black hole with higher mass or with spin of the extreme black hole will have higher time delay.
The effect of the charge of the black hole reduces the time delay between two relativistic images lying on the same side of the optical axis  when both travel  along either direct or retrograde  orbits. However, when one light ray is in the direct orbit but the other is in the retrograde orbit resulting in the images on the opposite side of the optical axis instead, the charge effect enhances the time delay between them. 
The estimate of the time delay due to galactic black holes and supermassive black holes ranges from less one hour to several days. We would like to emphasize that  the relation (\ref{ab_relation}) is found to be true in the Kerr and Kerr-Newman black holes that provide a more geometric interpretation to the time delay formulas we develop here, and also gives a handle given by $\vert b_{sc} \vert $ to estimate the time delay, namely $\Delta T \propto \vert b_{sc} \vert$  for nonextreme black holes. Together with the magnification of the relativistic images ($\mu \propto \vert b_{sc} \vert$) in \cite{Bozza_2003} as well as the angular position difference between two relativistic images ($\Delta \theta \propto \vert b_{sc} \vert$) in \cite{Hsieh_2021}, the relativistic images can be observed for large $\vert b_{sc} \vert$ leading to relatively long time delay, large  magnification and  angular separation between two neighboring images. According to \cite{Hsiao_2020} with the result of $b_{sc}$ as a function of $Q$, $a$ and also $M$ of the black holes, large mass $M$ for both direct and retrograde orbits  and angular momentum $a$ in the direct orbits enhance the observability of the images whereas large charge $Q$ for both direct and retrograde orbits and angular momentum $a$ in the retrograde orbits reduces the observability to see them.}

Our immediate next work will extend the current scope to study the trajectories of the light rays around the Kerr-Newman black holes on  nonequatorial plane by following \cite{Gralla_2020a,Gralla_2020b}.
Additionally, it will be of great interest to examine the relation between strong gravitational lensing and black hole quasinormal modes
for the Kerr and Kerr Newman black holes using the obtained the coefficients $\bar a$,$\bar b$ and $\tilde a$,$\tilde b$ due to the proposed idea by \cite{Stefanov_2010,Raffaelli_2016}.
%

\begin{acknowledgments}
This work was supported in part by the Ministry of Science and Technology, Taiwan, under Grant No.109-2112-M-259-003. We are so grateful to  the anonymous referee for the invaluable comments   with which to improve the quality of this paper.
We also thanks Chen-Yu Wang for  enlightening discussions.
\end{acknowledgments}

\section{appendix A}

We summarize the analytical expression of the smallest radius ${r_{sc}}$ of the circular motion of the light rays and the corresponding impact parameter $b_{sc}$. We also present here the SDL deflection angle in the form (\ref{hatalpha_as}) with the coefficients of $\bar a$ and $\bar b$
for the Kerr and Kerr-Newman black holes, respectively, from our previous paper \cite{Hsiao_2020,Hsieh_2021}.
\subsection{Kerr black holes}
The innermost trajectories of light rays  have
the smallest radius ${r_{sc}}$ given by
\begin{align} \label{rsc_k}
r_{sc}=  2 M\bigg\{ 1+ \cos\bigg[\frac{2}{3} \cos^{-1} \bigg( \frac{-sa}{M} \bigg) \bigg]\bigg\}
\;
\end{align}
obtained from the equation
\begin{align} \label{light_sphere_k}
  r_{sc}^2- 3 M r_{sc} + 2 a ( M r_{sc})^{1/2}=0 \,
\end{align}
with the corresponding impact parameter
\begin{align} \label{bsc_k}
 b_{sc} & =-a+ s6 M \cos\bigg[\frac{1}{3} \cos^{-1} \bigg( \frac{-sa}{M} \bigg) \bigg] \, .
 \end{align}
Combining (\ref{rsc_k}) and (\ref{bsc_k}) leads to
\begin{equation}
b_{sc}=a+\frac{r^2_{sc}}{\sqrt{M r_{sc}}}\; . \label{bsc_k2}
\end{equation}


The coefficients $\bar a$ and $\bar b$ in (\ref{hatalpha_as}) are
\begin{equation}
\begin{split} \label{abar_k_f}
\bar{a}=&\frac{r_{sc}^3}{\sqrt{c_{2 sc}
}}\left[ \frac{C_{-sc}}{r_{sc}r_{-}-a^2}+\frac{C_{+sc}}{r_{sc}r_{+}-a^2}\right]\;,
\end{split}
\end{equation}
\begin{equation}
\begin{split}\label{bbar_k}
\bar{b}=&-\pi+\bar{a} \log{\left( \frac{36}{7+4\sqrt{3}}\frac{8c_{2sc}^2 b_{sc}''}{c_{1sc}'^2 b_{sc}} \right)}\\
&+\frac{r_{sc}^3}{\sqrt{c_{2 sc}}}\frac{2 a C_{-sc}}{ a^2-r_{sc}r_- } \frac{\sqrt{3}}{\sqrt{a^2+2r_{sc}r_-}} \log{\left( \frac{\sqrt{a^2+2r_{sc}r_-}-a}{\sqrt{a^2+2r_{sc}r_-}+a} \frac{\sqrt{a^2+2r_{sc}r_-}+\sqrt{3}a}{\sqrt{a^2+2r_{sc}r_-}-\sqrt{3}a}\right)}\\
&+\frac{r_{sc}^3}{\sqrt{c_{2sc}}}\frac{2 a C_{+sc}}{ a^2-r_{sc}r_+ } \frac{\sqrt{3}}{\sqrt{a^2+2r_{sc}r_+}} \log{\left( \frac{\sqrt{a^2+2r_{sc}r_+}-a}{\sqrt{a^2+2r_{sc}r_+}+a} \frac{\sqrt{a^2+2r_{sc}r_+}+\sqrt{3}a}{\sqrt{a^2+2r_{sc}r_+}-\sqrt{3}a}\right)}\,,
\end{split}
\end{equation}
where $z_-$, $z_+$ are defined in (\ref{C_tilde_pm_k}), and the coefficients $C_-$, $C_+$ are
\begin{eqnarray}\label{Cpm_k}
C_-&=&\frac{a^2-2Mr_-(1-\frac{a}{b_s})}{2r_0 \sqrt{M^2-a^2}}\, ,\\
C_+&=&\frac{-a^2+2Mr_+(1-\frac{a}{b_s})}{2r_0 \sqrt{M^2-a^2}}
\end{eqnarray}
with  $r_+$ ($r_-$) being the outer (inner) horizon of a Kerr black hole defined in (\ref{rpm_k}) and $c_1, c_2,c_3$ are given by (\ref{c1_k})-(\ref{c3_k}).

\subsection{Kerr-Newman black holes}

As for the Kerr-Newman black hole, the solution of $r_{sc}$ of the radius of the innermost circular motion has been found in \cite{Hsiao_2020} as
\begin{align}\label{rsc_kn}
 r_{sc} & =\frac{3M}{2}+\frac{1}{2\sqrt{3}}\sqrt{9M^2-8Q^2+U_{c}+\frac{P_{c}}{U_{c}}} \no\\
 &\quad -\frac{s}{2}\sqrt{6M^2-\frac{16Q^2}{3}-\frac{1}{3}\left(U_c+\frac{P_{c}}{U_{c}}\right)
 +\frac{8\sqrt{3}Ma^2}{\sqrt{9M^2-8Q^2+U_c+\frac{P_c}{U_c}}}} \;\; ,
 \end{align}
where
 \begin{align}
 P_{c} & =(9M^2-8Q^2)^2-24a^2(3M^2-2Q^2) \, , \\
 U_{c} & =\bigg\{(9M^2-8Q^2)^3-36a^2(9M^2-8Q^2)(3M^2-2Q^2)+216M^2a^4 \no\\
 &\quad\quad +24\sqrt{3}a^2\sqrt{(M^2-a^2-Q^2)\left[Q^2(9M^2-8Q^2)^2-27M^4a^2\right]}\bigg\}^\frac{1}{3} \,
 \end{align}
 from solving the following equation
 \begin{align}\label{light_sphere_kn}
 2 Q^2 + r_c^2- 3 M r_c + 2 a ( M r_c-Q^2)^{1/2}=0 \, .
\end{align}
The analytical expression of the critical value of the impact parameter ${b_{sc} }$ can be written as a function of black hole's parameters \cite{Hsiao_2020},
\begin{align} \label{bsc_kn}
 b_{sc} & =-a+\frac{M^2a}{2(M^2-Q^2)}+\frac{s}{2\sqrt{3}(M^2-Q^2)}\Bigg[\sqrt{V+(M^2-Q^2)\left(U+\frac{P}{U}\right)} \no\\
 &
 +\sqrt{2V-(M^2-Q^2)\left(U+\frac{P}{U}\right)-\frac{s6\sqrt{3}M^2a\left[(M^2-Q^2)(9M^2-8Q^2)^2-M^4a^2\right]}
 {\sqrt{V+(M^2-Q^2)\left(U+\frac{P}{U}\right)}}}\Bigg]\;,
 \end{align}
where
 \begin{align}
 P & =(3M^2-4Q^2)\left[9(3M^2-4Q^2)^3+8Q^2(9M^2-8Q^2)^2-216M^4a^2\right] \, , \\
 U & =\bigg\{-\left[3(3M^2-2Q^2)^2-4Q^4\right]\left[9M^2(9M^2-8Q^2)^3-8\left[3(3M^2-2Q^2)^2-4Q^4\right]^2\right] \, \no\\
 &\qquad +108M^4a^2\left[9(3M^2-4Q^2)^3+4Q^2(9M^2-8Q^2)^2-54M^4a^2\right]\,  \no\\
 &\qquad +24\sqrt{3}M^2\sqrt{(M^2-a^2-Q^2)\left[Q^2(9M^2-8Q^2)^2-27M^4a^2\right]^3}\bigg\}^\frac{1}{3} \, ,\\
 V & =3M^4a^2+(M^2-Q^2)\left[6(3M^2-2Q^2)^2-8Q^4\right] \,.
 \end{align}
Corresponding to (\ref{bsc_k2}) we have in this case \cite{Hsiao_2020}
\begin{equation}
b_{sc}=a+\frac{r^2_{sc}}{\sqrt{M r_{sc}-Q^2}}\; . \label{bsc_kn2}
\end{equation}
The analytical expressions of the coefficients $\bar a$ and $\bar b$ in (\ref{hatalpha_as}) can be summarized below depending on the black hole parameters explicitly and implicitly through $r_{sc}$ and $b_{sc}$ as
\begin{equation}\label{abar_kn_f}
\begin{split}
\bar{a}=&\frac{r_{sc}^3}{\sqrt{c_{2sc}}}\left[ \frac{C_{-sc}}{r_{sc} r_--(a^2+Q^2)}+\frac{C_{+sc}}{r_{sc} r_{+}-(a^2+Q^2)}\right]
\end{split}
\end{equation}
and
\begin{equation}\label{bbar_kn}
\begin{split}
\bar{b}=&-\pi+\bar{a} \log{\left[\frac{36}{4(1-c_{4sc}/c_{2sc})^2+4\sqrt{3}(1-c_{4sc}/c_{2sc})^{3/2}+3(1-c_{4sc}/c_{2sc})} \frac{8 c_{2sc}^2 b_{sc}''}{c_{1sc}'^2 b_{sc}}\right]}\\
&+\frac{r_{sc}^3}{\sqrt{c_{2sc}}}\frac{2(a^2+Q^2) C_{-sc}}{(a^2+Q^2-r_{sc}r_-)}\frac{\sqrt{3}}{{P_-}}\\
&\quad\quad \times \log\left[ \frac{-r_{sc}r_-}{a^2+Q^2-r_{sc}r_-}  \frac{\left({P_-}+\sqrt{3} ({a^2+Q^2}) \right)^2-3 \left({a^2+Q^2}-{r_{sc} r_-} \right)^2 ({c_{4sc}}/{c_{2sc}})}{\left(P_-+(a^2+Q^2) (1-c_{4sc}/c_{2sc})^{1/2}\right)^2-3 {r^2_{sc} r^2_-}  ({c_{4sc}}/{c_{2sc}})}\right]\\
&+\frac{r_{sc}^3}{\sqrt{c_{2sc}}}\frac{2 [(a^2+Q^2)C_{+sc} +(a^2+Q^2-r_{sc}r_+)C_{Qsc}]}{(a^2+Q^2-r_{sc}r_+)}\frac{\sqrt{3}}{{P_+}}\\
&\quad\quad \times \log{\left[ \frac{-r_{sc}r_+}{a^2+Q^2-r_{sc}r_+} \frac{\left({P_+}+\sqrt{3} ({a^2+Q^2}) \right)^2-3 \left({a^2+Q^2}-{r_{sc} r_+} \right)^2 ({c_{4sc}}/{c_{2sc}})}{\left(P_++(a^2+Q^2) (1-c_{4sc}/c_{2sc})^{1/2}\right)^2-3 {r^2_{sc} r^2_+}  ({c_{4sc}}/{c_{2sc}})}\right]} \, .
\end{split}
\end{equation}
The coefficients $c_1, c_2, c_3,$ and $ c_4$ are listed in (\ref{c1_kn})-(\ref{c4_kn}).
In the equation above we also have defineded
\begin{equation}
\begin{split}
P^2_{\pm}&=(a^2+Q^2+2r_{sc}r_{\pm})(a^2+Q^2)-(a^2+Q^2+r_{sc}r_{\pm})(a^2+Q^2-r_{sc}r_{\pm})\frac{c_{4sc}}{c_{2sc}}\, .
\end{split}
\end{equation}
The corresponding coefficients $C_-$, $C_Q$, and $C_+$ in the Kerr-Newman case are
\begin{eqnarray} \label{CpmQ_kn}
C_-&=&\frac{a^2+Q^2-2Mr_-(1-\frac{a}{b_s})+\frac{Q^2r_-^2}{a^2+Q^2}(1-\frac{a}{b_s})}{2r_0\sqrt{M^2-a^2-Q^2}}\, ,\\
C_Q&=&\frac{Q^2}{r_0^2}\left(1-\frac{a}{b_s}\right)\, ,\\
C_+&=&\frac{a^2+Q^2-2Mr_-(1-\frac{a}{b_s})+\frac{Q^2}{r_{0}}(r_+-r_-)(1-\frac{a}{b_s})+Q^2(1-\frac{a}{b_s})}
{-2r_{0}\sqrt{M^2-a^2-Q^2}} \, .
\end{eqnarray}

\section{appendix B}

 The integrals of the function $f_D(z,r_0)$ in (\ref{f_D_k}) and   $f_R(z,r_0)=f(z,r_0)-f_D(z,r_0)$ in the Kerr cases are obtained analytically as
\begin{equation} \label{I_D_k}
\begin{split}
I_D(r_0)=&\int_0^Z f_D(z,r_0) dz\\
=&\frac{r_0^3}{a^2b}\frac{\tilde{C}_-}{\sqrt{c_1z_-+c_2z_-^2}}
\log{\left(\frac{\sqrt{c_1z_-+c_2z_- Z}+\sqrt{Z} \sqrt{c_1+c_2z_-}}
{\sqrt{c_1z_-+c_2z_- Z}-\sqrt{Z} \sqrt{c_1+c_2z_-}} \right)} \\
&+\frac{r_0^3}{a^2b}\frac{\tilde{C}_+}{\sqrt{c_1z_++c_2z_+^2}}
\log{\left(\frac{\sqrt{c_1z_++c_2z_+Z}+\sqrt{Z} \sqrt{c_1+c_2z_+}}
{\sqrt{c_1z_++c_2z_+ Z}-\sqrt{Z} \sqrt{c_1(+c_2z_+}} \right)} \\
&+\frac{r_0^3}{a^2b} \frac{r_0^2}{2 \sqrt{c_1+c_2 Z}}
\left[ -\frac{2\sqrt{Z}(c_1+c_2Z)}{(c_1+c_2)(Z-1)(z_--1)(z_+-1)} \right. \\
&-\frac{2\sqrt{c_1+c_2Z}}{(z_--1)^2 \sqrt{c_1z_-+c_2z_-^2}(z_--z_+)} \log{ \left( \frac{\sqrt{c_1z_-+c_2z_-Z}+\sqrt{c_1Z+c_2z_-Z}}{\sqrt{c_1z_-+c_2z_-Z}-\sqrt{c_1Z+c_2z_-Z}} \right) } \\
&+\frac{2\sqrt{c_1+c_2Z}}{(z_+-1)^2 \sqrt{c_1z_++c_2z_+^2}(z_--z_+)} \log{ \left( \frac{\sqrt{c_1z_++c_2z_+Z}+\sqrt{c_1Z+c_2z_+Z}}{\sqrt{c_1z_++c_2z_+Z}-\sqrt{c_1Z+c_2z_+Z}} \right) } \\
&+\frac{\sqrt{c_1+c_2Z}[c_1(5+z_-(z_+-3)-3z_+)+2c_2(3+z_-(z_+-2)-2z_+)]}{(c_1+c_2)^{3/2} (z_--1)^2(z_+-1)^2} \log{\left(\frac{1+\sqrt{Z}}{1-\sqrt{Z}}\right)} \\
&+\frac{\sqrt{c_1+c_2Z}[c_1(5+z_-(z_+-3)-3z_+)+2c_2(3+z_-(z_+-2)-2z_+)]}{(c_1+c_2)^{3/2} (z_--1)^2(z_+-1)^2} \\
&\times \left.\log{\left(\frac{c_1+c_2\sqrt{Z}+\sqrt{c_1+c_2} \sqrt{c_1+c_2Z}    }{ c_1-c_2\sqrt{Z}+\sqrt{c_1+c_2} \sqrt{c_1+c_2Z} }\right)} \right] \, .
\end{split}
\end{equation}

\begin{equation}
\begin{split} \label{I_R_k}
I_R(r_{0}) &=\int_0^Z f_R(z,r_{0}) dz\\
=&\frac{r_{0}}{b} \frac{r_{0}^2}{a^2} \frac{\tilde{C}_-}{\sqrt{c_2}z_-}\log{\left(\frac{z_-}{z_--Z} \frac{\sqrt{c_2+c_3Z}+\sqrt{c_2}}{\sqrt{c_2+c_3Z}-\sqrt{c_2}} \frac{c_3Z}{4c_2} \right)}\\
&+\frac{r_{0}}{b} \frac{r_{0}^2}{a^2}\frac{\tilde{C}_-}{\sqrt{c_2+c_3z_-}z_-}\log{\left( \frac{\sqrt{c_2+c_3z_-}-\sqrt{c_2+c_3Z}} {\sqrt{c_2+c_3z_-}+\sqrt{c_2+c_3Z}} \frac{\sqrt{c_2+c_3z_-}+\sqrt{c_2}}{\sqrt{c_2+c_3z_-}-\sqrt{c_2}}\right)}\\
&+\frac{r_{0}}{b} \frac{r_{0}^2}{a^2}\frac{\tilde{C}_+}{\sqrt{c_2}z_+}\log{\left(\frac{z_+}{z_+-Z} \frac{\sqrt{c_2+c_3Z}+\sqrt{c_2}}{\sqrt{c_2+c_3Z}-\sqrt{c_2}} \frac{c_3Z}{4c_2} \right)}\\
&+\frac{r_{0}}{b} \frac{r_{0}^2}{a^2}\frac{\tilde{C}_+}{\sqrt{c_2+c_3z_+}z_+}\log{\left( \frac{\sqrt{c_2+c_3z_+}-\sqrt{c_2+c_3Z}} {\sqrt{c_2+c_3z_+}+\sqrt{c_2+c_3Z}} \frac{\sqrt{c_2+c_3z_+}+\sqrt{c_2}}{\sqrt{c_2+c_3z_+}-\sqrt{c_2}}\right)}\\
&+\frac{r_{0}}{b} \frac{r_{0}^2}{a^2}
\left[  \frac{r_{sc}^2Z}{\sqrt{c_2}(Z-1)(z_--1)(z_+-1)} \right. \\
&-\frac{\sqrt{c_2}r_{0}^2}{(c_2+c_3)(z_--1)(z_+-1)}\\
&-\frac{\sqrt{c_2+c_3Z}r_{0}^2}{(c_2+c_3)(Z-1)(z_--1)(z_+-1)}\\
&+\frac{r_{0}^2[2c_2(3+z_-(z_+-2)-2z_+)+c_3(7-5z_++z_-(3z_+-5))]}{2(c_2+c_3)^{3/2} (z_--1)^2(z_+-1)^2} \\
&\times \log{\left( \frac{\sqrt{c_2+c_3}+\sqrt{c_2+c_3Z}} {\sqrt{c_2+c_3}-\sqrt{c_2+c_3Z}} \frac{\sqrt{c_2+c_3}-\sqrt{c_2}}{\sqrt{c_2+c_3}+\sqrt{c_2}}\right)}\\
&+\frac{r_{0}^2}{(z_--1)^2z_-\sqrt{c_2+c_3z_-}(z_--z_+)}\log{\left( \frac{\sqrt{c_2+c_3z_-}-\sqrt{c_2+c_3Z}} {\sqrt{c_2+c_3z_-}+\sqrt{c_2+c_3Z}} \frac{\sqrt{c_2+c_3z_-}+\sqrt{c_2}}{\sqrt{c_2+c_3z_-}-\sqrt{c_2}}\right)}\\
&+\frac{r_{0}^2}{(z_+-1)^2z_+\sqrt{c_2+c_3z_+}(-z_-+z_+)}\log{\left( \frac{\sqrt{c_2+c_3z_+}-\sqrt{c_2+c_3Z}} {\sqrt{c_2+c_3z_+}+\sqrt{c_2+c_3Z}} \frac{\sqrt{c_2+c_3z_+}+\sqrt{c_2}}{\sqrt{c_2+c_3z_+}-\sqrt{c_2}}\right)}\\
&-\frac{r_{0}^2[-3-z_-(z_+-2)+2z_+]}{\sqrt{c_2}(z_--1)^2(z_+-1)^2} \log{\left( \frac{c_2(1-Z)}{c_2+c_3Z} \right)}\\
&+\frac{r_{0}^2}{\sqrt{c_2}z_-z_+} \log{\left( \frac{\sqrt{c_2+c_3Z}-\sqrt{c_2}}{\sqrt{c_2+c_3Z}+\sqrt{c_2}} \frac{c_2+c_3Z}{4c_3Z} \right)}\\
&-\frac{r_{0}^2}{\sqrt{c_2}(z_--1)^2z_-(z_--z_+)} \log{\left( \frac{c_2}{c_2+c_3Z}\frac{z_--Z}{z_-} \right)}\\
&\left. -\frac{r_{0}^2}{\sqrt{c_2}(z_+-1)^2z_+(-z_-+z_+)} \log{\left( \frac{c_2}{c_2+c_3Z}\frac{z_+-Z}{z_+} \right)}  \right]    \;.
\end{split}
\end{equation}

The analytical expressions of $I_D$ and $I_R$ for the case of Kerr-Newman black holes are similar to the formulas given above, but lengthy, and are omitted in this paper.

\end{document}